\documentclass[sn-mathphys-ay]{sn-jnl}
\usepackage[english]{babel}
\usepackage[title]{appendix}
\usepackage{color}
\usepackage[table]{xcolor}%
\usepackage{colortbl}
\usepackage{textcomp}%
\usepackage{manyfoot}%
\usepackage{graphicx}
\usepackage[normal]{caption}
\usepackage{subcaption}
\usepackage{booktabs}
\usepackage{array}
\usepackage{float}
\usepackage{tikz}
\usetikzlibrary{arrows, decorations.pathmorphing, decorations.markings, backgrounds, positioning, fit, petri}
\usepackage{quantikz}
\usepackage{pdflscape}
\usepackage{multirow}%
\usepackage{comment}
\usepackage{algorithm2e}
\usepackage{pdfpages}

\usepackage{longtable}
\usepackage{lscape}
\usepackage{threeparttable}

\usepackage{amsmath,amsthm,amssymb,amsfonts}%
\usepackage{bm}
\usepackage{eucal}
\usepackage{braket,mleftright}
\usepackage[version=3]{mhchem}
\usepackage{siunitx}
\theoremstyle{plain}

\theoremstyle{plain}

\theoremstyle{plain}

\begin{document}

\title[Quantum Recovery Rate]{Hybrid Quantum Neural Networks with Amplitude Encoding for Credit Recovery Rate Prediction}


\author[1,2]{\fnm{Ying} \sur{Chen}}\email{matcheny@nus.edu.sg}

\author[3]{\fnm{Paul} \sur{Griffin}}\email{paulgriffin@smu.edu.sg}

\author*[4,6]{\fnm{Paolo} \sur{Recchia}}\email{paolo\_re@nus.edu.sg}

\author*[4]{\fnm{Lei} \sur{Zhou}}\email{leizhou@nus.edu.sg}

\author[5]{\fnm{Hongrui} \sur{Zhang}}\email{zhanghongrui@u.nus.edu}

\affil[1]{\orgdiv{Centre for Quantitative Finance, Department of Mathematics}, \orgname{National University of Singapore}, \orgaddress{\street{10 Lower Kent Ridge Rd, National University of Singapore}, \city{Singapore}, \postcode{119076}, \state{Singapore}}}

\affil[2]{\orgdiv{Risk Management Institute}, \orgname{National University of Singapore}, \orgaddress{\street{11 Kent Ridge Dr, National University of Singapore}, \city{Singapore}, \postcode{119244}, \state{Singapore}}}

\affil[3]{\orgdiv{School of Computing and Information Systems}, \orgname{Singapore Management University}, \orgaddress{\street{80 Stamford Rd}, \city{Singapore}, \postcode{178902}, \state{Singapore}}}

\affil[4]{\orgdiv{Asian Institute of Digital Finance}, \orgname{National University of Singapore}, \orgaddress{\street{Innovation 4, Research Link}, \city{Singapore}, \postcode{117602}, \state{Singapore}}}

\affil[6]{\orgdiv{Centre for Quantum Technologies}, \orgname{National University of Singapore}, \orgaddress{\street{3 Science Drive 2, Block S15}, \city{Singapore}, \postcode{117543}, \state{Singapore}}}

\affil[5]{\orgdiv{Department of Mathematics}, \orgname{National University of Singapore}, \orgaddress{\street{10 Lower Kent Ridge Rd, National University of Singapore}, \city{Singapore}, \postcode{119076}, \state{Singapore}}}


\abstract{
Recovery rate prediction plays a pivotal role in bond investment strategies by enhancing risk assessment, optimizing portfolio allocation, improving pricing accuracy, and supporting effective credit risk management. However, accurate forecasting remains challenging due to complex nonlinear dependencies, high-dimensional feature spaces, and limited sample sizes—conditions under which classical machine learning models are prone to overfitting. We propose a hybrid Quantum Machine Learning (QML) model with Amplitude Encoding, leveraging the unitarity constraint of Parametrized Quantum Circuits (PQC) and the exponential data compression capability of qubits.  We evaluate the model on a global recovery rate dataset comprising 1,725 observations and 256 features from 1996 to 2023. Our hybrid method significantly outperforms both classical neural networks and QML models using Angle Encoding, achieving a lower Root Mean Squared Error (RMSE) of 0.228, compared to 0.246 and 0.242, respectively. It also performs competitively with ensemble tree methods such as XGBoost. While practical implementation challenges remain for Noisy Intermediate-Scale Quantum (NISQ) hardware, our quantum simulation and preliminary results on noisy simulators demonstrate the promise of hybrid quantum-classical architectures in enhancing the accuracy and robustness of recovery rate forecasting. These findings illustrate the potential of quantum machine learning in shaping the future of credit risk prediction.}





\maketitle

\section{Introduction}
\label{sec:intro}

Recovery rate prediction is an essential component of credit risk management, considered alongside key metrics such as Exposure at Default (EAD) and Probability of Default (PD)~\cite{BCBS2023}. While EAD and PD assess the likelihood and magnitude of credit losses, recovery rates measure the proportion of funds that can be recovered following a default. Recovery rates are particularly important for risk assessment, portfolio optimization, and pricing strategies.
Despite their importance, recovery rates have received relatively little attention in practice. A common simplifying assumption is to use a constant recovery rate, typically around 40\%, despite empirical evidence showing a wide range from 0\% to 100\%, often with bimodality around 10\% and 100\%~\cite{Pykthin2003, Andersen2004, Berd2005, Gambetti2018}. This oversimplification can result in inaccurate risk assessments, suboptimal investment decisions, and flawed pricing models—particularly for distressed or lower-rated bonds. This limited attention is partly due to the technical challenges involved in accurately modeling recovery rates, including data sparsity, high-dimensional feature spaces, and complex nonlinear dependencies.

There are technical challenges in forecasting recovery rates. Traditional statistical models—such as linear regression, logistic regression, and generalized linear models—have shown limited effectiveness in recovery rate prediction due to their inability to model the highly nonlinear and interactive effects among firm characteristics, macroeconomic indicators, and debt contract features. As highlighted by~\cite{acharya2007creditor, qi2011comparison, nazemi2018macroeconomic}, these models are particularly sensitive to outliers, which are prevalent in distressed debt data, and tend to perform poorly when the underlying data distributions are skewed or exhibit heavy tails—common features in recovery rate datasets. Moreover, their reliance on fixed functional forms and strong distributional assumptions (e.g., normality, homoscedasticity) limits their flexibility in capturing heterogeneity across different sectors or credit environments. These models also scale poorly in high-dimensional settings, where the number of explanatory variables is large relative to the sample size, making them prone to overfitting and unstable parameter estimates. 

Classical machine learning approaches have addressed training stability and generalization issues by constraining weight matrices to orthogonal or unitary forms, using techniques such as Orthogonal Neural Networks (OrthNNs) and Unitary Neural Networks (UNNs)~\cite{bartlett1996valid, le2015simple, henaff2016recurrent, bansal2018can, lezcano2019cheap, li2019orthogonal, wang2020orthogonal, mashhadi2021parallel}. These constraints ensure that each layer acts as a local isometry due to the bounded spectral norm of orthogonal and unitary matrices.~\cite{li2019orthogonal} showed that optimal generalization is achieved when weight matrices have equal singular values. This reduces sensitivity to perturbations and serves as an implicit regularizer~\cite{bartlett2017spectrally, li2019orthogonal}. Theoretically, bounding the spectral norm also lowers capacity measures such as Rademacher complexity and fat-shattering dimension~\cite{bartlett1996valid, bartlett2017spectrally}. Empirically, orthogonality has been shown to improve test performance and robustness across tasks~\cite{bansal2018can, cisse2017parseval, wang2020orthogonal}. However, enforcing orthogonality during gradient-based training is computationally demanding, often requiring re-orthogonalization steps with time complexity $O(N^3)$ for input size $N$~\cite{li2019orthogonal}. This underscores the need for novel approaches.

The computational challenges in recovery rate prediction stem from two intertwined factors: the inherently complex relationships between financial variables (e.g., non-linear dependencies between macroeconomic indicators and borrower defaults), the small-sample nature of recovery rate data (e.g., limited historical default records for niche markets) which demands a specialized approach to prevent overfitting. Quantum Neural Networks (QNNs) present a promising alternative by leveraging key properties of quantum systems: the exponential growth of Hilbert space with the number of qubits enables QNNs to represent high-dimensional features with fewer parameters, while the intrinsic unitarity of quantum circuits naturally constrains model capacity and helps reduce the risk of overfitting. For instance,~\cite{havlicek2019supervised} demonstrated that a quantum-enhanced kernel method achieves 98\% accuracy on a 100-image MNIST subset, outperforming classical SVM (92\%), by utilizing entanglement and superposition to efficiently extract patterns from minimal data. Furthermore,~\cite{schuld2019quantum} theoretically established that QNNs exhibit slower growth of Rademacher complexity with increasing qubits, suggesting superior generalization in low-data scenarios. These unique advantages make QNNs a compelling alternative for prediction problems that combine data scarcity with high-dimensional feature spaces. Motivated by this, our study investigates the feasibility and performance of QNNs in the context of recovery rate prediction—an application that not only poses significant challenges for classical approaches, but also offers an ideal testbed for evaluating the generalization potential of quantum models under realistic constraints.

A critical aspect of Quantum Neural Networks (QNNs) is the encoding of classical data into quantum states. A recent study by~\cite{schetakis2024quantum}, through the combination of a QNN with a classical neural network, built a Quantum Machine Learning (QML) model for credit scoring of Small and Medium Enterprises (SMEs). The proposed QML model achieved performance comparable to classical models with fewer training epochs. Their model employed Angle Encoding, which maps classical inputs to rotation angles of single-qubit gates~\cite{schuld2018supervised, ranga2024quantum, rath2024quantum, gong2024quantum}. While this method simplifies circuit design and enables shallow circuits well-suited to current Noisy Intermediate-Scale Quantum (NISQ) hardware, its scalability is limited because the number of required qubits grows linearly with the input dimensionality. As a result, it becomes inefficient for high-dimensional datasets. In contrast, our study adopts Amplitude Encoding, which encodes data into the amplitudes of a quantum state, allowing for exponentially more compact representations~\cite{schuld2018supervised, benedetti2019parameterized, ranga2024quantum, rath2024quantum}. The required number of qubits grows only logarithmically with the input size. This leads to significantly improved scalability, especially for datasets with large feature spaces. Moreover, using fewer qubits results in fewer trainable parameters within the Parameterized Quantum Circuit (PQC), thereby enhancing computational efficiency.

To sum up, in this paper, we propose a Quantum Machine Learning (QML) model that leverages Amplitude Encoding to directly address key challenges in recovery rate prediction. Unlike prior work that primarily uses Angle Encoding~\cite[e.g.,][]{schuld2020circuit, mitarai2018quantum}, our approach efficiently encodes high-dimensional financial data into quantum amplitudes, achieving exponential input compression. By integrating amplitude-encoded QNN with classical neural layers, our model achieves parameter efficiency without compromising expressiveness. This hybrid architecture effectively addresses three central challenges: small-sample learning, computational complexity, and relational nonlinearity—challenges that are difficult to overcome using existing classical or quantum methods alone.

The proposed method demonstrated superior performance using a global dataset of 1,725 observations with 256 features spanning 576 firms from 1996 to 2023. It achieved a Root Mean Square Error (RMSE) of 0.228, lower than the RMSE of 0.246 for classical Neural Networks and 0.242 for quantum models with Angle Encoding (see the Results section below for more details). Additionally, the QML model with Amplitude Encoding has fewer trainable parameters, leading to a faster training time of 0.73 seconds per epoch compared to the 0.81 seconds necessary for the QML with Angle Encoding. The lower qubit requirements and reduced computation time underscore the practical applicability of our method for recovery rate forecasting. 

This paper makes three main contributions to the literature on QML and financial risk modeling. First, unlike the widely used Angle Encoding, we propose a QML model that uses Amplitude Encoding, more suitable for high-dimensional data representation with fewer qubits~\cite{schuld2014quest, zoufal2019quantum}. In contrast to prior QML studies such as \cite{schetakis2024quantum}, which primarily evaluate quantum architectures on low-dimensional synthetic datasets, and \cite{schuld2020circuit}, which focus on classification tasks with relatively simple structures, our setting requires modeling a high-dimensional (256-feature) heterogeneous financial dataset with a continuous recovery-rate target. To address these domain-specific characteristics (i.e., high dimension, samll size and complex relationships), we design an architecture in which Amplitude Encoding serves as a principled dimensionality-reduction mechanism, mapping the 256-dimensional input vector to an 8-qubit quantum state while preserving geometric relationships between observations through unitary normalization. This property makes the encoding particularly well-suited for credit-risk data, where normalization and stability are crucial for mitigating the effects of volatility clustering and scale dispersion.
Our work contributes to ongoing research on how different data encoding strategies impact the scalability and practical applicability of quantum models~\cite{abbas2021power, kyriienko2022unsupervised}.

Second, we apply the QML to the task of predicting recovery rates—an important topic in credit risk modeling. Previous studies have mostly relied on linear models or classical machine learning techniques~\cite{altman2005link, qi2011comparison, Nazemi2018improving}. In our empirical analysis, the QML with Amplitude Encoding performs better than both traditional models and QML with Angle Encoding. 
The QML with Amplitude Encoding delivers higher prediction accuracy, more stable results across different runs, and faster convergence, especially when the training data are limited and the relationship between variables is nonlinear.

Third, our approach offers a general modeling tool for financial problems that involve small datasets, high-dimensional features, and complex variable interactions. These conditions are common in many areas of financial risk management and often limit the effectiveness of standard models~\cite{bertsimas2016best, bellotti2012loss}. By showing that the QML with Amplitude Encoding can work well in such settings, we contribute a new method for analyzing difficult financial prediction problems where data are scarce and patterns are complex.

\color{black}
The remainder of this paper is organized as follows: 
Section~\ref{sec:literature} reviews related literature on recovery rate modeling and Quantum Machine Learning (QML).
Section~\ref{sec:data} presents the data and its features. Section~\ref{sec:meth} details the hybrid QML approach and Amplitude Encoding. 
Section~\ref{sec:results} discusses the numerical analysis, and Section~\ref{sec:concl} concludes with insights and directions for future research.

\section{Literature Review}\label{sec:literature}
Understanding the determinants of recovery rates given default is critical for credit risk modeling, pricing, and risk management. While a vast literature has emerged on modeling the probability of default (PD) (see, e.g., \cite{duffie2009frailty}; \cite{duan2012multiperiod}), the modeling of recovery rates has received relatively limited attention until recently. Yet, recovery rates play a central role in determining the expected loss and are essential for accurate credit risk assessments and financial regulations.

This literature review organizes the related studies into four major streams: (i) conventional econometric modeling techniques; (ii) advanced machine learning methods;  (iii) methodological challenges and innovations in high-dimensional recovery modeling and (iv) Quantum Machine Learning applications. We also highlight key differences across these approaches and identify how recent developments contribute to addressing gaps in the literature.

\subsection{Traditional Approaches to Recovery Rate Modeling}
Traditional recovery rate modeling has primarily relied on econometric techniques that assume linear or parametric relationships between recovery outcomes and explanatory variables. These approaches include linear regression, Tobit models, ordered choice models, and structural credit risk models, each with specific assumptions and limitations.

Linear regression models represent the earliest and most straightforward approach.~\cite{altman2005link} apply this method using bond data, modeling recovery rates as a function of debt seniority, collateral, and macroeconomic conditions.~\cite{acharya2007creditor} similarly employ regression analysis to study how governance structures and creditor rights affect recoveries. These models are easy to implement and interpret but fail to address the bounded nature of recovery rates and ignore potential non-linearities.

To address boundary issues, some studies adopt Tobit models. Since recovery rates are censored between 0 and 1, Tobit models are suitable for dealing with observations that pile up at these extremes.~\cite{bellotti2012loss} offer a comprehensive review of LGD models, advocating for Tobit-based approaches when dealing with censored recovery data.~\cite{qi2011comparison} use Tobit specifications to assess the impact of loan and borrower characteristics on recovery outcomes in U.S. bank loans.

Another common method is the use of ordered logit or probit models, which discretize recovery rates into categories.~\cite{calabrese2014predicting} models recovery as an ordered outcome to better handle its bimodal distribution.~\cite{dermine2006bank} adopt similar techniques when analyzing recovery rates of European bank loans, emphasizing loan type and provisioning practices.

Beyond these reduced-form approaches, structural credit risk models offer a more theoretical perspective. These models derive recovery as an endogenous outcome of firm value at the time of default.~\cite{duan2012public} develop a structural model in which recovery depends on firm leverage and the timing of default.~\cite{giesecke2011corporate} extend the Merton framework by integrating dynamic firm fundamentals and market conditions.~\cite{duan2014predicting} implement a structural framework that emphasizes the importance of credit cycle effects in explaining recovery rate dynamics.

These approaches often suffer from limitations in flexibility and scalability. The parametric structure requires pre-specification of functional forms, and the small number of predictors leaves out potentially important nonlinear effects and interaction terms. This has led to a growing interest in semi-parametric or non-parametric methods that can better accommodate the complex structure of the recovery process.

However, these methods still face significant challenges when dealing with high-dimensional predictor spaces. Additionally, they often require strong parametric assumptions and may not generalize well across datasets with small sample sizes—a persistent issue in recovery rate research due to the rarity of default events.

In summary, traditional models offer important insights into recovery processes but suffer from rigidity in functional form, limited capacity to capture complex interactions, and challenges in accommodating high-dimensional data. These limitations motivate the use of more flexible and scalable machine learning methods, as explored in this project.

\subsection{Machine Learning Methods in Recovery Modeling}
Recent years have witnessed growing interest in Machine Learning (ML) methods for recovery modeling, driven by the increasing availability of high-dimensional and unstructured data, as well as the limitations of traditional approaches. ML models are particularly well-suited to capture nonlinear relationships, complex interactions, and heterogeneity in the data without requiring pre-specified functional forms.

Support Vector Machines (SVMs) and k-Nearest Neighbor (k-NN) classifiers have also been applied in the credit risk domain, although they are less common in recovery modeling specifically. ML models often outperform traditional approaches in terms of predictive accuracy but face challenges related to overfitting, interpretability, and the curse of dimensionality.

Some recent work explores text-based features, such as sentiment extracted from financial news or filings, to improve prediction performance. For instance, \cite{araci2019finbert} and \cite{nassirtoussi2015text, nassirtoussi2014text} apply NLP techniques in credit scoring, and similar ideas are adapted in this project to model recovery using risk sentiment indices derived from news.

Tree-based ensemble methods such as random forests and gradient boosted trees (e.g., XGBoost) have been used to model recovery rates due to their ability to handle variable interactions and nonlinearity~\citep{hastie2009elements, nazemi2022machine, liu2024machine}. 

Neural networks and deep learning techniques~\cite{goodfellow2016deep} offer even greater modeling flexibility, although their application is limited by data sparsity in credit markets. In practice, the small sample size of defaults and limited labeled data on recovery rates make the application of deep learning challenging in credit markets.

We provide an overview of key studies modeling recovery rates for U.S. corporate bonds in Table~\ref{tab:literature}. Existing literature has adopted a wide range of methodologies, which have evolved significantly over time.
\begin{table}[htbp]
    \centering
    \footnotesize
    \setlength{\tabcolsep}{5.5pt}
    \renewcommand{\arraystretch}{1.05}
    \resizebox{\columnwidth}{!}{
    \begin{tabular}{lll}
        \toprule
        Study & Methodology & Model Type (Key Technique) \\
        \midrule
        \cite{altman1996defaulted} & Linear Regression & Traditional econometric \\
        \cite{tibshirani1996regression} & LASSO & Regularization \\
        \cite{hamilton2001default} & Linear Regression & Traditional econometric \\
        \cite{frye2000loss} & Conditional model & Traditional econometric \\
        \cite{franks2004recovery} & Probit / Macro analysis & Traditional \\
        \cite{varma2005determinants} & Univariate and multivariate regression & Traditional \\
        \cite{yuan2006model} & Group LASSO & Structured Regularization \\
        \cite{acharya2007creditor} & Regression & Traditional econometric \\
        \cite{hastie2009elements} & Random Forests / Boosting & ML (tree-based) \\
        \cite{qi2011comparison} & Quantile Regression & Semi-parametric \\
        \cite{jacobs2011modeling} & Beta-link generalized linear model & Semi-parametric \\
        \cite{bellotti2012loss} & Tobit & Traditional \\
        \cite{friewald2012illiquidity} & Regression with market variables & Traditional econometric \\
        \cite{jankowitsch2014recovery} & Zero-One Inflated Beta & Semi-parametric \\
        \cite{calabrese2014predicting} & Ordered Logit & Traditional \\
        \cite{duan2014predicting} & Structural Credit Risk Model & Traditional with macro factors \\
        \cite{nassirtoussi2015text} & Text Mining + ML & Machine Learning (NLP) \\
        \cite{altman2014mixture} & Parametric regressions, regression trees & Ensemble + Traditional \\
        \cite{donovan2015accounting} & Univariate and multivariate regression & Traditional \\
        \cite{mora2015determinants} & Univariate and multivariate regression & Traditional \\
        \cite{bertsimas2016best} & Best Subset Selection & Regularization / Optimization \\
        \cite{goodfellow2016deep} & Neural Networks & Deep Learning \\
        \cite{kim2016determinants} & Multivariate linear regression & Traditional \\
        \cite{lundberg2017unified} & SHAP values & Model interpretation \\
        \cite{kalotay2017intertemporal} & Inverse Gaussian, regression trees, mixtures & Semi-parametric / ML \\
        \cite{nazemi2018macroeconomic} & SVRs, bagging, boosting, LASSO, ridge & Machine Learning \\
        \cite{araci2019finbert} & NLP-based Sentiment & Machine Learning (BERT) \\
        \cite{gambetti2019modeling} & Beta regression, mixture models, regression trees & Semi-parametric + ML \\
        \bottomrule
    \end{tabular}%
    }
    \caption{Summary of Recovery Rate Modeling Approaches in the Literature.}
    \label{tab:literature}
\end{table}

\subsection{Quantum Machine Learning}
Quantum Machine Learning (QML) models combine Quantum Neural Networks (QNNs) with classical layers. This new emerging field, leveraging the power of quantum computing with Machine Learning techniques, addresses problems commonly challenging for classical algorithms. Thanks to quantum parallelism, QML models have attracted attention for their potential to process high-dimensional data and capture complex nonlinearities in a more compact architecture. QNNs leverage quantum properties, including superposition, entanglement, and unitary transformations, to encode and transform input data via a Parameterized Quantum Circuit (PQC). These models are especially promising in scenarios where classical models suffer from the curse of dimensionality, such as in high-dimensional recovery rate modeling with sparse default data.

Recent studies have explored the application of QML in finance and credit risk. For example,~\cite{schuld2014quest} and~\cite{biamonte2017quantum} provide foundational discussions on how quantum algorithms can be adapted to supervised and unsupervised learning tasks. More recent work by~\cite{abbas2021power} and~\cite{benedetti2019parameterized} explores how QNNs can approximate functions in complex spaces and perform optimization tasks over rugged objective surfaces. These features make QNNs especially suitable for credit risk modeling, where the signal-to-noise ratio is low and default-related data is sparse. In practice, quantum-enhanced models such as Variational Quantum Circuits (VQCs), Quantum Boltzmann Machines (QBM), and Quantum Convolutional Neural Networks (QCNNs) have shown success in tasks like classification and regression with limited data~\cite{alcazar2020classical} and success in market forecasting, where Quantum Elman Neural Networks have proven effective for sequential data tasks~\cite{liu2022quantum}. Similarly,~\cite{havlicek2019supervised} shows that quantum models can perform competitively on structured prediction tasks and kernel-based learning in domains with limited training observations. Additionally, QML models have shown advantages in time series forecasting when implemented on Quantum Processing Units (QPUs)~\citep{emmanoulopoulos2022quantum, rivera2022time}. The Quantum Amplitude Estimation (QAE) algorithm~\cite{brassard2002quantum}, known for its quadratic speedup over classical Monte Carlo techniques~\cite{montanaro2015quantum}, presents a promising alternative for probabilistic modeling tasks, such as those encountered in option pricing. Building on the foundations of QAE, other QML approaches, including Quantum Generative Adversarial Networks (qGANs)~\cite{zoufal2019quantum}, have gained traction for probabilistic modeling, demonstrating their effectiveness in applications like option pricing. Furthermore,~\cite{plekhanov2022variational} introduced Variational Quantum Amplitude Estimation (VQAE), which combines classical variational optimization with QAE. This approach ensures that the circuit depth remains below a desired threshold, highlighting its potential for practical applications in financial pricing tasks. In fraud detection, QNNs have also demonstrated superior performance, achieving better precision and lower false-positive rates compared to classical methods~\cite{kyriienko2022unsupervised, tekkali2023smart}. 

As quantum hardware continues to evolve, QNNs offer a pathway to overcoming data limitations and complexity in financial modeling. In this paper, we adopt a QML model to capture complex nonlinear interactions in high-dimensional feature spaces and to enhance the recovery rate prediction under data-scarce conditions. By leveraging hybrid quantum-classical architectures, our approach aims to exploit the representational power of QML while maintaining computational tractability on near-term quantum devices.




\color{black}
\section{Data}
\label{sec:data}

We consider a dataset comprising 256 features and 1,725 observations covering 576 firms from 1996 to 2023. The data is obtained through the NRF Research Project UP5 of the National University of Singapore. The UP5 data contains Macroeconomic and market-related features obtained from FRED and Refinitiv; financial statement features of firms sourced from Bloomberg; and bond-level features, as well as firm-level or market-level credit product features provided by the Credit Research Initiative (CRI) of the National University of Singapore. 

The recovery amount of each bond in this study is defined as the bond's price 30 days after the default date. This is the most common way used in the literature.~\cite{Moodys2011} uses a price “roughly” 30 days after the default event. Early S\&P reports use the average price 30 to 45 days post-default, while more recent S\&P reports focus on exactly 30 days afterward.~\cite{jankowitsch2014determinants} use average prices of the first 30 default days. We follow the literature and use the 30-day period. 

\subsection{Summary Description of Recovery Rate}
\begin{figure}
        \centering
        \includegraphics[scale=.6]{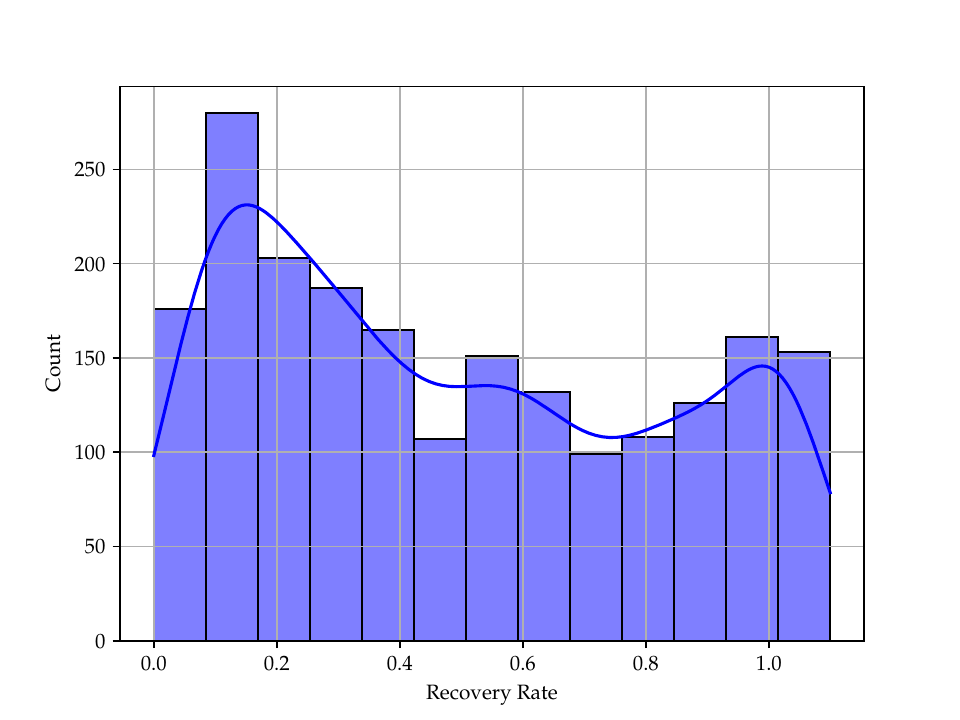}
         \caption{This histogram with a kernel density estimate (smooth blue line) shows the distribution of recovery rates ranging from 0 to 1.1 for defaulted bonds.}
         \label{recovery rate}
\end{figure}

Table~\ref{tab:recovery_stats} summarizes the characteristics of recovery rates. It has a mean value of 48\%, a median value of 42\%, and a standard deviation (STD) of 0.33. Figure~\ref{recovery rate} displays the histogram of recovery rates. In the Figure, the y-axis represents the frequency of each bar, whereas the x-axis the recovery rate. It reveals a broad range of recovery rates. They are almost all distributed between 0 and 1 and occasionally exceed 1.  The histogram also exhibits a bimodal pattern with primary and secondary peaks around 10\% and 100\% respectively. Although recovery rates tend to cluster around 40$\%$ \footnote{
   \cite{jankowitsch2014determinants} find an average recovery rate of 38.6\% for 2002-2010, while The average ultimate recovery rate for US corporate bonds reported by~\cite{cantor2007moody} is 37\% for defaults between 1987 and 2006. In general, market participants tend to assume constant recovery rates of around 40\% within the pricing models~\cite{das2009implied}.}, 
they exhibit significant variability, with a large standard deviation. This wide distribution highlights the limitations of assuming a fixed recovery rate (such as 40$\%$) in pricing models, which oversimplifies the complex and dynamic nature of actual recovery rates and can be misleading. Assuming a fixed recovery rate leads to inaccurate risk assessments, flawed pricing strategies, and miscalculated credit risk metrics. 

\begin{table}[htbp]
    \centering
    \begin{tabular}{lcccccccc}
        \toprule
        & Obs. & Mean & STD & Min & 25\% & 50\% & 75\% & Max \\
        \midrule
        Recovery Rate & 1,725 & 0.4845 & 0.3317 & 0 & 0.1811 & 0.4170 & 0.7896 & 1.0996 \\
        \bottomrule
    \end{tabular}
    \caption{Summary Statistics of Recovery Rate}
    \label{tab:recovery_stats}
\end{table}

The relationship between various features and recovery rates of defaulted bonds is highly complex and nonlinear, which makes traditional linear models inadequate for accurate predictions. 
Previous studies have highlighted the intricate interactions between different factors influencing recovery rates. For example,~\cite{altman1996defaulted} finds that industry-specific characteristics, such as public utilities and chemicals, influence recovery rates significantly.~\cite{altman2005link} note that macroeconomic variables like GDP growth and stock market returns have weak correlations with recovery rates, while factors like default rates, seniority, and collateral levels play a more direct role.~\cite{acharya2007creditor} further documents that recovery rates are lower in distressed industries, emphasizing the importance of industry-specific dynamics.
Additionally, models based on mixtures of Gaussian distributions, as introduced by~\cite{altman2014mixture}, show superior out-of-time forecasting accuracy compared to traditional parametric models. Similarly, nonparametric approaches like regression trees and support vector machines, as demonstrated by~\cite{yao2017enhancing} and~\cite{nazemi2018macroeconomic}, outperform linear regression in terms of prediction accuracy, especially in out-of-sample scenarios. These studies indicate that nonlinear relationships, including interactions between bond characteristics, market conditions, and macroeconomic factors, are better captured by more flexible machine learning models. Thus, a neural network capable of modeling such complex and nonlinear relationships is well-suited for predicting recovery rates of defaulted bonds.

\begin{figure}
    \includegraphics[scale=0.4]{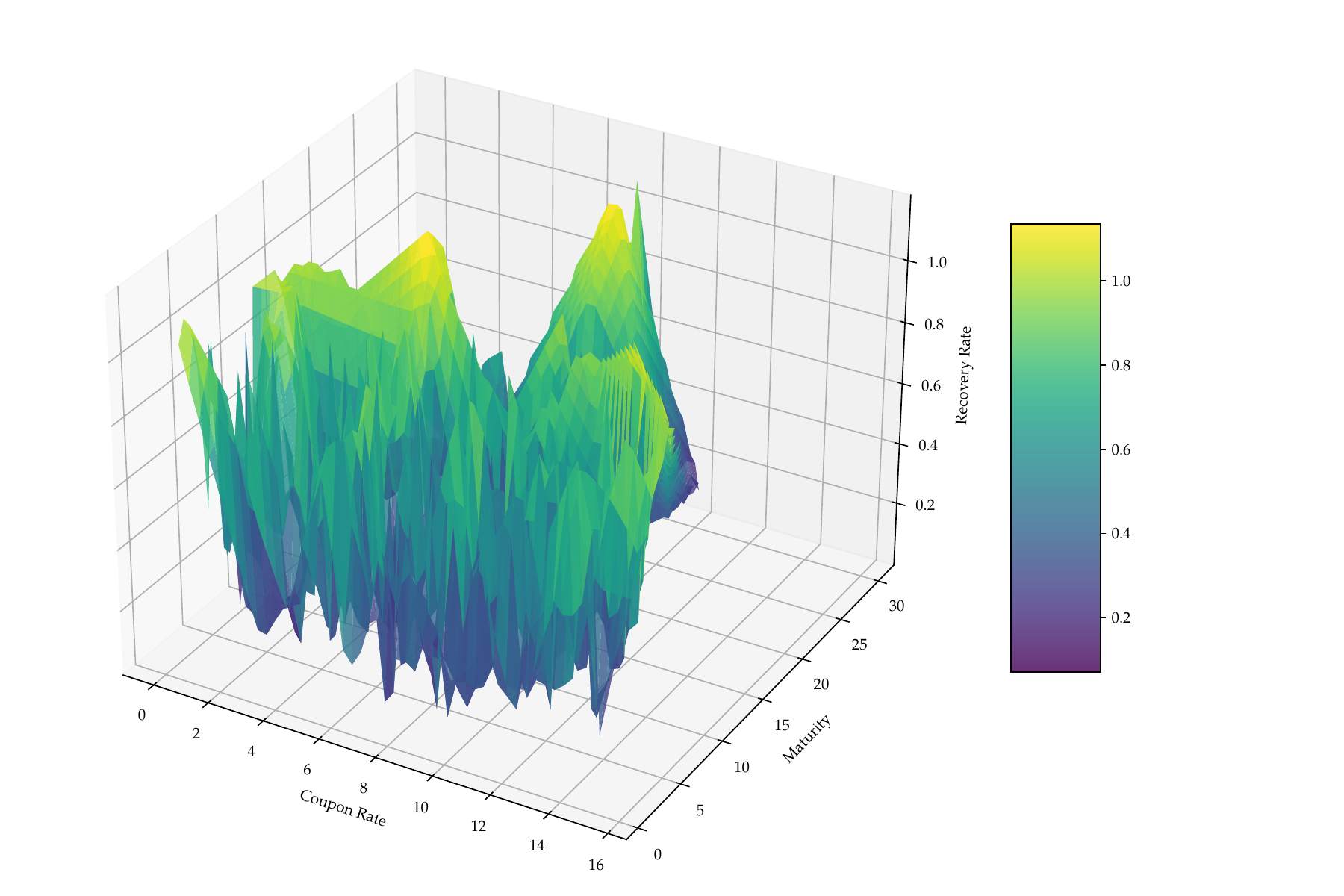}
    \caption{
    The plot visualizes how the recovery rate varies with changes in the coupon rate and maturity. The surface is generated through cubic interpolation to provide a smooth representation of the underlying trend. The color gradient of the surface indicates the magnitude of the recovery rate, with darker shades corresponding to lower recovery rates and lighter shades indicating higher recovery rates.}
    \label{fig:rr_3d}
\end{figure}

\begin{figure}
    \centering
    \includegraphics[scale=.6]{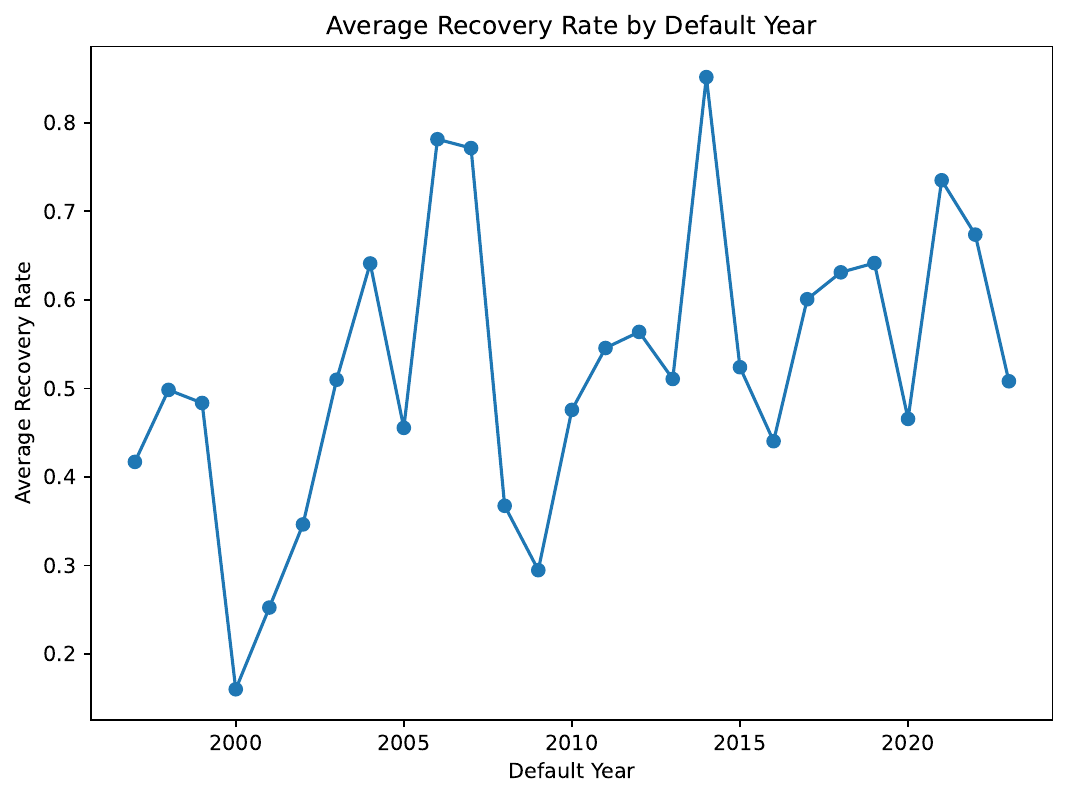}
    \caption{
    The plot illustrates the trend of recovery rates over time, with each data point representing the average recovery rate for a given default year. The x-axis shows the default year, while the y-axis represents the corresponding average recovery rate. }
    \label{fig:rr_ts}
\end{figure}

\begin{figure}
    \centering
    \includegraphics[scale=0.6]{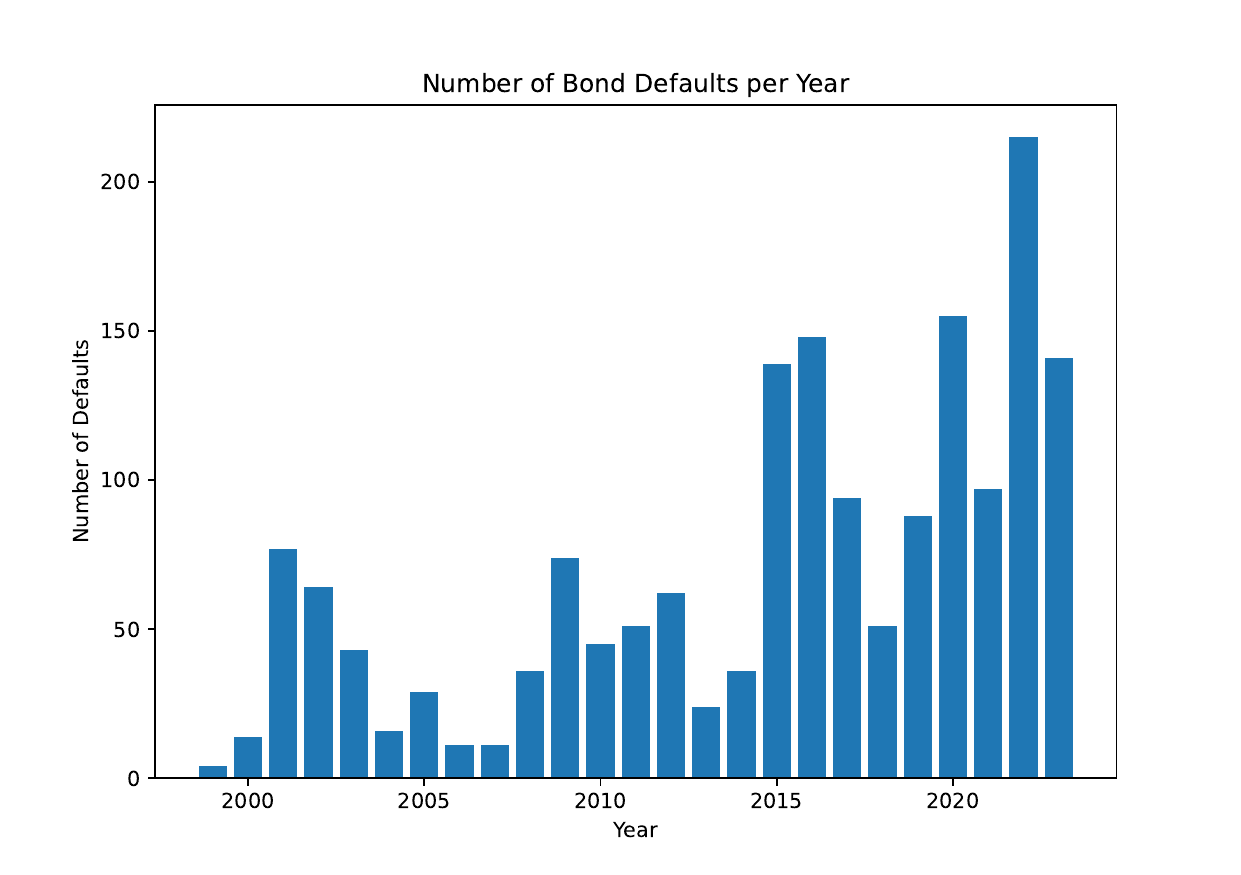}
    \caption{ The plot illustrates yearly counts of defaulted bonds in the sample from 1999 to 2023. The x-axis shows the default year, while the y-axis represents the corresponding counts of recovery rate. 
    }
    \label{fig:default_counts}
\end{figure}

Our own data also reveals the complex nature of recovery rate prediction. For instance, when we examine the relationship between recovery rate, coupon rate, and maturity, we observe no clear linear relationship. A 3D plot of the recovery rate against these features, as shown in Figure~\ref{fig:rr_3d}, implies that the recovery rate is influenced by multiple factors in non-linear ways, further emphasizing the inadequacy of traditional linear regression models. Additionally, we observe considerable variability in average recovery rates across different years in Figure~\ref{fig:rr_ts}, which reflects the impact of changing market conditions. Figure~\ref{fig:default_counts} plots the yearly distribution of defaulted bonds in our sample. A notable feature of the dataset is the substantial imbalance across years. Defaults are relatively sparse in the early part of the sample, with a total of 597 defaulted bonds between 1999 and 2015. In contrast, defaults become considerably more frequent after 2015, with 1,128 observations from 2016 to 2023. This sharp increase reflects well-known credit cycle dynamics, including post-crisis restructuring waves and sector-specific downturns during the late 2010s and early 2020s. 

These patterns suggest that forecasting recovery rates using traditional statistical regression models would be ineffective. Thus, we adopt neural networks, which are better equipped to handle the nonlinearity and complexity of the data.

\subsection{Overview of Features}

In total, we construct 256 features that encompass firm-level characteristics, bond-specific attributes, and macroeconomic indicators. These features are selected based on their relevance to credit risk and recovery rate modeling, as supported by prior literature such as~\cite{jankowitsch2014determinants} and~\cite{gambetti2022meta}. For ease of presentation and interpretation, we organize these features into several broad categories as shown in Table~\ref{tab:feature_categories} \footnote{The complete list of all 256 features is provided in the supplementary materials.}.

\begin{table}[htbp]
    \centering
    \begin{tabular}{p{0.35\linewidth}p{0.5\linewidth}p{0.1\linewidth}}
        \toprule
        Category & Examples & Count \\
        \midrule
        Bond-Specific Variables & Coupon rate, duration, seniority, default event type & 29 \\
        Firm-level Characteristics & Return on assets (ROA), debt-to-equity ratio, current ratio & 17 \\
        Credit Risk Measures & Probability of default (PD) in 1 month, distance to default (DTD), sigma & 78 \\
        Industry, Geographic, and Currency Variables & Industry dummies, region identifiers, currency indicators & 98 \\
        Macroeconomic and Market Conditions & GDP growth, inflation, financial uncertainty indices, treasury yields & 34 \\
        \bottomrule
    \end{tabular}
    \caption{Summary of feature categories used in recovery rate prediction. The table groups the 256 features into five major categories and provides illustrative examples for each type.}
    \label{tab:feature_categories}
\end{table}

The first group contains bond-specific variables. These include information that is intrinsic to the bond contract itself, such as the coupon rate, duration, and seniority level. We also include a series of event indicators capturing credit events like defaults, restructurings, delistings, and insolvencies. In total, this category comprises 29 variables. The data for this group are primarily obtained from Bloomberg and Refinitiv.

The second group focuses on firm-level financial indicators. These variables are constructed using company financial statements and include standard measures such as return on assets (ROA), debt-to-equity ratio, current ratio, and market capitalization. These metrics provide insight into the issuing firm’s solvency and operating health. This category includes 17 variables and is sourced entirely from Bloomberg.

The third category consists of credit risk measures, particularly probability of default (PD) estimates. We incorporate a comprehensive range of PD scores calculated across different aggregation levels, including sector, domicile, exchange, and global scopes. These variables are forward-looking and help capture variation in credit quality across firms. In total, we include 77 PD-related variables, all obtained from the Credit Research Initiative (CRI) of the National University of Singapore. 
In addition to the PD measures, this category also includes one volatility-based variable $\sigma$ derived from daily equity returns, which provides an additional dimension of firm-level credit risk. In total, the credit risk category therefore contains 78 variables.

The fourth group includes categorical variables that account for structural heterogeneity across firms. Specifically, we introduce one-hot encoded dummies for industries, regions, and currencies. These variables allow us to control for country-specific and sector-specific effects. There are 55 industry dummies and 43 regional or currency indicators, summing to 98 variables in this category. These are also provided by the CRI.

Finally, we include a group of macroeconomic and market-related variables. This group captures the broader economic environment and includes variables such as interest rates of various maturities, GDP growth, inflation, financial uncertainty indices, and measures of market volatility. These 34 variables are collected from FRED and Bloomberg, and they serve to contextualize firm and bond behavior within the prevailing macro-financial landscape.

All features are selected based on domain expertise and prior empirical findings that link these attributes to credit outcomes and recovery dynamics. 

Before training, all continuous input features are standardized using z-score normalization, computed from the training set only. This prevents scale imbalances across heterogeneous financial variables and avoids any form of temporal or cross-fold information leakage. In contrast, following common practice in time-series and tabular forecasting, the categorical True/False features are simply mapped to 0 and 1, preserving their discrete nature and preventing distortions that could arise from z-scaling a two-point distribution. The selected dataset contains no missing entries in the variables retained for modeling, so no imputation step is required. We also do not apply principal component analysis (PCA) or other dimensionality-reduction techniques. Our architecture already includes an initial classical preprocessing layer that learns a compact representation jointly with the quantum circuit, making PCA unnecessary for this application.

\color{black}
\section{Methodology}
\label{sec:meth}

Our proposed Quantum Machine Learning (QML) model adopts a hybrid architecture, consisting of a classical preprocessing layer followed by a Quantum Neural Network (QNN) (the overall architecture is depicted in Figure~\ref{fig:ampl}). This design is inspired by~\cite{schetakis2024quantum}, who demonstrated the effectiveness of integrating classical neural components with quantum circuits in financial classification task. The overall pipeline of our hybrid QML model comprises the following four steps:
\begin{enumerate}
    \item \textbf{Classical Preprocessing (pink block in Figure~\ref{fig:ampl}):} The raw input features ($N=2^n$ dimensions) are first processed by a classical neural network, which acts as a feature transformation or embedding layer to condition the data for quantum encoding.

    \item \textbf{Quantum State Preparation (connection arrow between the pink block and the dashed block inside the purple block in Figure~\ref{fig:ampl}, and Figure~\ref{fig:ampl_enc}):} The transformed features are encoded into a quantum state $\ket{\psi}$ via Amplitude Encoding. This encoding scheme efficiently compresses high-dimensional data, requiring only $n=\log_2 N$ qubits for $N$ features.  

    \item \textbf{Parametrized Quantum Circuit (the solid block inside the purple block in Figure~\ref{fig:ampl}):} The encoded quantum state is then passed through a Parameterized Quantum Circuit (PQC), composed of rotation gates and CNOT gates which produce entanglement.

    \item \textbf{Measurement and Output:} After quantum processing, the qubits are measured in the computational basis. The measurement results are forwarded to a classical output layer to generate the final prediction (i.e., the recovery rate).
\end{enumerate}

While the execution of QML models on real quantum hardware remains constrained by noise, decoherence, and limited qubit counts, this study is conducted using a fault-tolerant quantum simulator. This allows us to isolate and evaluate the algorithmic performance of the proposed architecture without the confounding effects of hardware-induced errors.

\begin{figure}
\resizebox{1.\linewidth}{!}{
    \begin{tikzpicture}
        \node[draw, thick, rectangle, fill=red!30] (box1) at (0,0) {
            \begin{tikzpicture}[
            neuron/.style={circle, draw, minimum size=0.8cm}, 
            layer/.style={draw=none,fill=none},
            ->, >=stealth',
            dashedline/.style={draw=black, dashed}
            ]
    
            \node[neuron] (I1) at (0, 3) {};
            \node[neuron] (I2) at (0, 2) {};
            \node[neuron] (I3) at (0, 1) {};
            \node at (0, 0.2) {\vdots};
            \node[neuron] (I4) at (0, -0.7) {};
            \node[layer] at (0, -1.3) {$2^n$ features};
        
            \node[neuron] (I21) at (1.5, 3) {};
            \node[neuron] (I22) at (1.5, 2) {};
            \node[neuron] (I23) at (1.5, 1) {};
            \node at (1.5, 0.2) {\vdots};
            \node[neuron] (I24) at (1.5, -0.7) {};
            \node[layer] at (1.5, 3.6) {$2^n$ Act.};
        
            \draw[->] (I1) -- (I21);
            \draw[->] (I1) -- (I22);
            \draw[->] (I1) -- (I23);
            \draw[->] (I1) -- (I24);
            \draw[->] (I2) -- (I21);
            \draw[->] (I2) -- (I22);
            \draw[->] (I2) -- (I23);
            \draw[->] (I2) -- (I24);
            \draw[->] (I3) -- (I21);
            \draw[->] (I3) -- (I22);
            \draw[->] (I3) -- (I23);
            \draw[->] (I3) -- (I24);
            \draw[->] (I4) -- (I21);
            \draw[->] (I4) -- (I22);
            \draw[->] (I4) -- (I23);
            \draw[->] (I4) -- (I24);
            \end{tikzpicture}
        };
        \node[above=5pt of box1] {$2^n$ Input Classical Data};
        
        \node[draw, thick, rectangle, fill=blue!30] (box2) at (7.0, 0) {
            \begin{quantikz}
                \lstick{} & \gate[3]{\ket{\psi}} \gategroup[3,steps=1,style={dashed, inner sep=6pt},label style={label position=above,yshift=0.2cm}]{Amplitude Enc.} & \qw & \gate{R(\alpha_1, \beta_1, \gamma_1)} \gategroup[3,steps=5,style={inner sep=6pt},label style={label position=above,yshift=0.2cm}]{Strongly Entangling PQC.}   & \ctrl{1} & \qw & \qw & \targ{} & \meter{Z} & \lstick{} \\
                \lstick{} &  & \qw & \gate{R(\alpha_2, \beta_2, \gamma_2)} & \targ{} & \ctrl{1} & \qw & \qw & \meter{Z} & \lstick{}\\
                \lstick{} &  & \qw & \gate{R(\alpha_3, \beta_3, \gamma_3)} & \qw & \targ{} & \qw & \ctrl{-2} & \meter{Z} & \lstick{}
            \end{quantikz}
        };
        \node[above=5pt of box2] {QNN {\bf Amplitude Encoding}};
    
        \node (out) at (12.7,0) {
            \begin{tikzpicture}[
                neuron/.style={circle, draw, minimum size=0.8cm},
                dropout/.style={circle, draw, minimum size=0.8cm, dashed},
                layer/.style={draw=none,fill=none}
            ]
                \node[neuron] (O1) at (0, 0) {};
            \end{tikzpicture}
        };
        \node[above=5pt of out] {Output};
            
        \draw[->] (box1) -- (box2);
        \draw[->] (box2) -- (out);
    \end{tikzpicture}
    }
    \caption{The QML model with Amplitude Encoding. We encode a set of $N = 2^n$ classical data into the amplitude of the input quantum state denoted as $\ket{\psi}$. After the application of the Strongly Entangling PQC, a measurement is performed. These measurement results are then sent to the classical output layer and post-processed in the classical optimizer.}
    \label{fig:ampl}
\end{figure}

\subsection{Classical Preprocessing }
As shown in Figure~\ref{fig:ampl}, the input classical data—comprising $N=2^n$ features—is first processed by a classical neural network (pink block). The classical master layer comprises an input layer and a hidden layer of equal size, using a LeakyReLU activation function. This architecture offers three advantages. First, it extracts meaningful internal representations from high-dimensional, redundant input data. Second, it acts as a feature transformation or embedding layer to prepare the data for quantum data encoding. Third, it introduces non-linearity, enabling the model to capture complex relationships.

\subsection{Quantum Data Encoding}

To interface between the classical and quantum layers, we encode the transformed classical data into a quantum state, denoted as $\ket{\psi}$ (see Figures~\ref{fig:ampl} and~\ref{fig:ampl_enc}). 

Two commonly used methods for quantum data encoding are Angle Encoding and Amplitude Encoding, illustrated in Figures~\ref{fig:ang_enc} and~\ref{fig:ampl_enc}, respectively. 

Angle Encoding~\cite{schuld2018supervised, ranga2024quantum, rath2024quantum, gong2024quantum} maps classical data onto the rotation angles of single-qubit gates. While straightforward, this method requires one qubit per input feature, making it computationally demanding and impractical for high-dimensional datasets like those used in our recovery rate prediction due to the limited availability of logical qubits in current quantum hardware and classical simulators.

To overcome this limitation, we adopt Amplitude Encoding, which maps a $2^n$-dimensional classical feature vector into the amplitude components of an $n$-qubit quantum state as follows
\begin{equation}
\label{eq:orig_qs_a}
    \ket{\psi} = \sum_{i=0}^{N-1} \alpha_i \ket{b_1^i \dots b_n^i},
\end{equation}
where $N = 2^n$ is the number of features, $b_1^i, \dots, b_n^i$ is the binary representation of integer $i$, and $\alpha_i$ are the normalized amplitudes derived from the input data.

\begin{figure}[htbp]
    \begin{subfigure}{0.4\textwidth}
        \centering

    \begin{tikzpicture}
    \node (qc) at (0, 0) {
        \begin{quantikz}
        \lstick{} & \push{\ket{0}}  & \gate{R_x(x_1)}  & \lstick{} & \qw \\
        \lstick{} & \push{\ket{0}}  & \gate{R_x(x_2)}  & \lstick{} & \qw \\
        \lstick{} & \push{\ket{0}}  & \gate{R_x(x_3)}  & \lstick{} & \qw \\
        \lstick{} & \push{\ket{0}}  & \gate{R_x(x_4)}  & \lstick{} & \qw
        \end{quantikz}
    };
    \begin{pgfonlayer}{background}
        \useasboundingbox (qc.north west) rectangle (qc.south east);
        \node[draw, thick, dashed, fill=blue!30, fit=(qc)] {};
    \end{pgfonlayer}

\end{tikzpicture}
        \caption{Angle Encoding.}
        \label{fig:ang_enc}
    \end{subfigure} 
    \begin{subfigure}{0.55\textwidth}
        \centering
        \begin{tikzpicture}
        \node (qc) at (0, 0) 
        {
\begin{quantikz}
            \lstick{} & \push{\ket{0}} &  \gate[2]{\ket{\psi(X)}} & \qw  \\
            \lstick{} & \push{\ket{0}}  &  & \qw
        \end{quantikz}
        };
        \begin{pgfonlayer}{background}
            \useasboundingbox (qc.north west) rectangle (qc.south east);
            \node[draw, thick, dashed, fill=blue!30, fit=(qc)] {};
        \end{pgfonlayer}
        \end{tikzpicture}
        \caption{Amplitude Encoding.}
        \label{fig:ampl_enc}
    \end{subfigure}
    \caption{Two common approaches to encode an example of four classical data features $X = x_1,x_2,x_3,x_4$. In Figure~\ref{fig:ang_enc}, the four classical features are mapped into the rotation angles of the four one-qubit $R_x$ rotation gate (Angle Encoding). In Figure~\ref{fig:ampl_enc}, the features are mapped into the amplitude of the two-qubits state $\ket{\psi}$ (Amplitude Encoding). To prepare this quantum state, one-qubit $R_y(\bm \theta)$ rotation gates and CNOT gates must be applied, where the angles $\bm \theta = \theta_1,\theta_2,\theta_3$ depend on the four classical data $X = x_1,x_2,x_3,x_4$ (see~\cite{mottonen2004quantum} or the supplementary material provided).}
    \label{fig:diff_enc}
\end{figure}

The preparation of the state $\ket{\psi}$ in Amplitude Encoding follows the protocol described in~\cite{mottonen2004transformation}. This method uses a sequence of single-qubit $R_y(\bm{\theta})$ rotations and CNOT gates to transform the initial zero state $\ket{\bm{0}}$ into the desired amplitude-encoded state $\ket{\psi}$. 
As illustrated in Figure~\ref{fig:diff_enc}, for a four-dimensional input vector $X = \{x_1, x_2, x_3, x_4\}$, Angle Encoding requires four qubits (one per feature), while Amplitude Encoding only requires two qubits. This exponential compression enables us to scale to higher-dimensional datasets using fewer quantum resources. In our case, 256 input features are encoded using only eight qubits.
This figure essentially depicts the interface between the classical layer (the pink block in Figure~\ref{fig:ampl}) and the quantum neural network (the purple block in Figure~\ref{fig:ampl}). A full description of the Möttönen protocol is beyond the scope of this paper; however, we provide supplementary material highlighting the most relevant aspects of the Amplitude Encoding procedure for interested readers.

\subsection{Parametrize Quantum Circuit (PQC)}

Once encoded, the quantum state $\ket{\psi}$ is passed through a Parameterized Quantum Circuit (PQC) in the solid block of the QNN in Figure~\ref{fig:ampl}. The PQC consists of a sequence of parameterized single-qubit rotation gates and, to increase expressibility—the capacity to represent diverse quantum states and explore the Hilbert space~\cite{sim2019expressibility}—entangling CNOT gates. A typical PQC architecture includes:
\begin{enumerate}
    \item A layer of $n$ single-qubit rotation gates, $R(\alpha, \beta, \gamma)$, applied to each qubit. Here, $\alpha, \beta, \gamma$ are the trainable parameters.
    \item A layer of $n$ two-qubit entangling gates. Specifically, we use CNOT gates with a range of one: the $i$th qubit acts as the control for the $(i+1)$th target qubit.
\end{enumerate}
This configuration, commonly referred to as a Strongly Entangling PQC~\cite{schuld2020circuit}, provides the PQC with entangling power while maintaining a manageable number of parameters, specifically $O(3n)$ trainable parameters for $n$ qubits. The overall architecture results in a smaller number of trainable parameters in the QNN, specifically scaling as $O(3 \log_2 N)$, where $N$ represents the number of features. 

After the quantum computation, the expectation values of $Z$ Pauli observables are measured and passed to the classical output layer. The resulting measurement outcomes are then passed back to the classical domain, where they are post-processed by a final classical output layer for regression (i.e., recovery rate prediction). 

This hybrid setup effectively leverages the strengths of both classical and quantum components: the classical neural layer captures coarse patterns and performs feature conditioning, while the QNN with Amplitude Encoding provides a compact and expressive mechanism for modeling nonlinear dependencies within a constrained qubit budget.

\subsection{Unitarity and Generalization}

Quantum circuits used in the hybrid architecture consist entirely of unitary transformations, meaning that at the quantum state level, every operation preserves the norm of the state vector. In other words, the learned quantum feature map is always 1-Lipschitz continuous with respect to its quantum state inputs. More formally, for any unitary $U$ and any quantum states $\ket{\psi}, \ket{\phi}$
\begin{equation}
\label{eq:unit_const}
    \|U\ket{\psi} - U\ket{\phi}\| =  \|\ket{\psi} - \ket{\phi}\|.
\end{equation}
This ensures that the quantum feature transformations cannot arbitrarily stretch or distort the encoded data. This constraint has important implications for stability and generalization, and closely parallels results established for classical deep networks with orthogonal or spectrally-normalized weight matrices. In the Orthogonal Deep Neural Networks (OrthNNs) framework,~\cite{li2019orthogonal} show that the generalization error of a deep model can be bounded in terms of how closely each layer behaves as an isometry; the tightest bound arises when all singular values equal one, with orthogonal (or unitary) matrices being the optimal case. 

Similarly,~\cite{bartlett2017spectrally} derive a margin-based generalization bound that depends on the product of spectral norms of the weight matrices. Networks whose layers are constrained to have spectral norm one exhibit tighter control of the Lipschitz constant and therefore reduced effective capacity. PQCs naturally satisfy this regime: the quantum layers contribute a product of spectral norms equal to one, implying that the complexity of the hybrid model is dominated by the classical output head, while the quantum feature extractor itself is fully spectrally normalized.

Furthermore, unitarity also stabilizes the backpropagation of gradients through the quantum circuit. During training, the loss function $\mathcal{L}$ produces a gradient with respect to the circuit’s output features. Let
\begin{equation}
g = \nabla_f \mathcal{L}(f(x,\theta)),
\end{equation}
denote the gradient signal arriving at the quantum layer from subsequent classical layers. In backpropagation, this gradient is transformed by the adjoint of the circuit unitary, $U^\dagger(\theta)$. Since every unitary satisfies $\|U^\dagger v\| = \|v\|$ for all vectors $v$, we obtain
\begin{equation}
\label{eq:unit_g_const}
\|U^\dagger g\| = \|g\|.    
\end{equation}
This means the quantum layer cannot amplify the gradient norm, eliminating the risk of gradient explosion, an issue that commonly affects deep classical networks with unconstrained weight matrices \cite{hardt2016train}. Unitarity thus enforces a form of intrinsic gradient norm control, contributing to stable optimization and improved generalization in small-sample settings.

It is important to underline that the constraints expressed in Equations~\eqref{eq:unit_const} and~\eqref{eq:unit_g_const} do not act directly on classical vectors, but only on the internal quantum state of the PQC. However, although unitarity does not directly constrain the classical inputs, it strongly constrains the internal quantum representation and ensures bounded, stable gradients. This creates a meaningful inductive bias: the hybrid quantum architecture cannot represent overly complex or noise-sensitive transformations, which in practice promotes better generalization and convergence in data-scarce environments, such as in the recovery rate prediction.

\color{black}
\subsection{Alternative regression models for comparison evaluation}

To thoroughly evaluate the performance of our proposed QML model, we compare it against both classical and quantum alternatives. 
\begin{itemize}
    \item[] {\bf QML with Angle Encoding:} The first natural baseline is a QML model using Angle Encoding, inspired by~\cite{schetakis2024quantum}. To address the challenge posed by qubit limitations, we follow the approach outlined in~\cite{schetakis2024quantum}, incorporating a classical preprocessing layer (referred to as the auxiliary layer hereafter) to reduce the dimensionality of the input data. This auxiliary layer extracts key features and ensures that the number of variables aligns with the number of available qubits. The outputs of this classical layer are then encoded into the quantum circuit using Angle Encoding as in Figure~\ref{fig:ang_enc}. This architecture, illustrated in Figure~\ref{fig:angle}, enables scalability by allowing flexibility in selecting the number of qubits while maintaining the model’s expressiveness. The number of trainable parameters in the additional auxiliary layer and the QNN scales as $O(3Nn)$, where $N$ is the number of input features, and $n$ is the number of selected qubits.
    
    {\bf Fully Connected Neural Network (FNN):} Among classical models, the most direct counterpart for comparison is a Feedforward Neural Network (FNN), as our QML architecture closely resembles an FNN with a QNN replacing the hidden layer.

    
    \item[] {\bf XGBoost:} In our comparative analysis, we also include a baseline XGBoost~\cite{chen2016xgboost} regressor. Although this model does not mirror the architecture of our QML model, we include it due to its widespread adoption and strong empirical performance. 
\end{itemize}


Table~\ref{tab:all_mods} summarizes the models considered in this study, along with the subsequent sections where the comparison analysis is presented. Additionally, to disentangle the effect of orthogonal regularization from the contribution of the quantum layer, we consider an orthogonality-constrained neural network, in which the hidden-layer weight matrix is enforced to remain orthogonal during training. This model serves as a classical control baseline. The corresponding benchmark comparisons between the proposed QML model and the orthogonality-constrained NN are reported in Section 3 of the Supplementary Materials.

\begin{table}[htbp]
    \centering
    \begin{tabular}{l l}
    \toprule
    Model & Section  \\
    \midrule
    FNN--QML Amplitude Encoding &~\ref{sec:results_fnn} \\
    QML Angle Encoding--QML Amplitude Encoding &~\ref{sec:results_fnn} \\
    XGBoost--QML Amplitude Encoding &~\ref{sec:results_xgb} \\
    \bottomrule
    \end{tabular}
    \caption{The list of regression models considered for the comparison evaluation.}
    \label{tab:all_mods}
\end{table}



\begin{figure}
\resizebox{1.\linewidth}{!}{
\begin{tikzpicture}

    \node[draw, thick, rectangle, fill=red!30] (box1) at (0,0) {
    \begin{tikzpicture}[
    neuron/.style={circle, draw, minimum size=0.8cm}, 
    layer/.style={draw=none,fill=none},
    ->, >=stealth',
    dashedline/.style={draw=black, dashed}
    ]

    \node[neuron] (I1) at (0, 3) {};
    \node[neuron] (I2) at (0, 2) {};
    \node[neuron] (I3) at (0, 1) {};
    \node at (0, 0.2) {\vdots};
    \node[neuron] (I4) at (0, -0.7) {};
    \node[layer] at (0, -1.3) {$N$ features};

    \node[neuron] (I21) at (1.5, 3) {};
    \node[neuron] (I22) at (1.5, 2) {};
    \node[neuron] (I23) at (1.5, 1) {};
    \node at (1.5, 0.2) {\vdots};
    \node[neuron] (I24) at (1.5, -0.7) {};
    \node[layer] at (1.5, 3.6) {$N$ Act.};

    \node[neuron] (H1) at (3, 2.5) {};
    \node[neuron] (H2) at (3, 1.5) {};
    \node[neuron] (H3) at (3, 0.5) {};
    \node[layer] at (3, -0.4) {$n$ Auxiliary};

    \draw[->] (I1) -- (I21);
    \draw[->] (I1) -- (I22);
    \draw[->] (I1) -- (I23);
    \draw[->] (I1) -- (I24);
    \draw[->] (I2) -- (I21);
    \draw[->] (I2) -- (I22);
    \draw[->] (I2) -- (I23);
    \draw[->] (I2) -- (I24);
    \draw[->] (I3) -- (I21);
    \draw[->] (I3) -- (I22);
    \draw[->] (I3) -- (I23);
    \draw[->] (I3) -- (I24);
    \draw[->] (I4) -- (I21);
    \draw[->] (I4) -- (I22);
    \draw[->] (I4) -- (I23);
    \draw[->] (I4) -- (I24);

    \draw[->] (I21) -- (H1);
    \draw[->] (I21) -- (H2);
    \draw[->] (I21) -- (H3);
    \draw[->] (I22) -- (H1);
    \draw[->] (I22) -- (H2);
    \draw[->] (I22) -- (H3);
    \draw[->] (I23) -- (H1);
    \draw[->] (I23) -- (H2);
    \draw[->] (I23) -- (H3);
    \draw[->] (I24) -- (H1);
    \draw[->] (I24) -- (H2);
    \draw[->] (I24) -- (H3);
    \end{tikzpicture}
    };
    \node[above=5pt of box1] {Classical Layer};
    
    \node[draw, thick, rectangle, fill=blue!30] (box2) at (8.0, 0.35) {
        \begin{quantikz}
            \lstick{} & \gate{R_x(x_1)}	 \gategroup[3,steps=1,style={dashed, inner
sep=6pt}, label style={label position=below, yshift=-0.4cm}]{Angle Enc.} &\qw & \gate{R(\alpha_1, \beta_1, \gamma_1)}  \gategroup[3,steps=5,style={inner
sep=6pt}, label style={label position=below, yshift=-0.4cm}]{Strongly Entangling PQC.}& \ctrl{1} & \qw & \qw & \targ{} & \meter{Z} & \lstick{} \\
            \lstick{} & \gate{R_x(x_2)} & \qw & \gate{R(\alpha_2, \beta_2, \gamma_2)} & \targ{} & \ctrl{1} & \qw & \qw & \meter{Z} & \lstick{}\\
            \lstick{} & \gate{R_x(x_3)} & \qw & \gate{R(\alpha_3, \beta_3, \gamma_3)} & \qw & \targ{} & \qw & \ctrl{-2} & \meter{Z} & \lstick{}
        \end{quantikz}
        };
        \node[above=5pt of box2] {QNN {\bf Angle Encoding}};

    \node (out) at (14.0
    ,0.35) {
        \begin{tikzpicture}[
            neuron/.style={circle, draw, minimum size=0.8cm},
            dropout/.style={circle, draw, minimum size=0.8cm, dashed},
            layer/.style={draw=none,fill=none}
        ]
            \node[neuron] (O1) at (0, 0) {};
        \end{tikzpicture}
        };
        \node[above=5pt of out] {Output};
        
    \draw[->] (1.9,1.4) -- (3.2,1.65);
    \draw[->] (1.9,0.4) -- (3.2,0.55);
    \draw[->] (1.9,-0.6) -- (3.2,-0.55);
    \draw[->] (box2) -- (out);
\end{tikzpicture}
}
    
    \caption{The QML model with Angle Encoding. Starting from a set of $N$ input features, we introduce an auxiliary (classical) layer that reduces the number of inputs to match the number of qubits $n$ used in the quantum circuit. For visualization purposes, the figure shows an example with three qubits, but the number of qubits can be adjusted as needed. The outputs of the auxiliary layer are classical values, which are encoded into quantum states using Angle Encoding via single-qubit rotation gates of the form $ R_x(x_i)$. After applying a Strongly Entangling Parameterized Quantum Circuit (PQC), a measurement is performed. The resulting measurement outcomes are then passed to a classical output layer and further processed by a classical optimizer.}
    \label{fig:angle}
\end{figure}

\color{black}
\section{Results}
\label{sec:results}

In this section, we present the results produced by our models, based on the methodologies and settings outlined in Sections~\ref{sec:meth} and Table~\ref{tab:all_mods}. These findings offer insights into the predictive performance of both classical and QML approaches on the chosen dataset. The evaluation of the models is conducted using standard metrics, such as RMSE calculated through four-fold cross-validation, leave-one-out cross-validation, along with statistical significance tests to determine which model performs significantly better than the others. \footnote{
The strong temporal heterogeneity as shown in Figure~\ref{fig:default_counts} implies that a strict chronological split (e.g., train on 1996--2015 and test on 2016--2023) would leave an insufficient number of defaults for model estimation. Training the proposed hybrid quantum and classical baselines requires an adequate minimum sample size to avoid severe underfitting. For this reason, we adopt a four-fold cross-validation design with fold-specific preprocessing to prevent temporal leakage while ensuring sufficient data for model training.}

\subsection{Parameter Settings and Experimental Setup}
\label{sec:comp}

The parameters in the QML and FNN regression models are optimized to minimize RMSE on the training data selected in a cross-validation setting. The optimization process utilized the Adam optimizer~\cite{kingma2014adam}, with specific hyperparameters such as learning rate and batch size detailed in Table~\ref{tab:specs}. These hyperparameters were selected through a systematic grid search to ensure optimal model performance. Table~\ref{tab:specs} also lists the computational resources used for training and evaluation, including both classical and quantum setups.

Deep learning practitioners commonly utilize neural network models optimized with the Adam optimizer. These models have well-established, high-performing implementations across various frameworks. For our implementation, we employ Python (version 3.12.3) and the PyTorch framework (version 2.4.1) with CUDA (version 12.1) to enable GPU acceleration. All experiments presented in the following sections were conducted on an NVIDIA GeForce RTX™ 4050 Laptop GPU.

The QML models were implemented using PennyLane (version 0.38.0), an open-source framework for quantum programming and differentiable PQCs. All QML models with PQCs were executed on a state-vector quantum simulator. Specifically, the built-in PennyLane device called \texttt{default.qubit}~\cite{pennylane_plugins,pennylane_defaultQ}. The \texttt{default.qubit} device, written in Python with Autograd and PyTorch backends, simulates quantum operations and performs measurements on quantum systems using a classical CPU. In particular, we conducted our quantum experiments on the 13th Gen Intel® Core™ i9-13900H CPU.

It is worth noting that alternative quantum devices could be employed~\cite{pennylane_plugins}, including those with GPU acceleration, tensor network implementations, or density matrix simulators for noisy environments. However, our experiments focused on fault-tolerant PQCs with a limited number of qubits (no more than 14). In this context, the \texttt{default.qubit} state-vector simulator proved to be the most efficient and effective choice.

Regarding gradient computation for PQC parameters, all quantum experiments were performed on a classical computer, allowing gradient calculation via automatic differentiation and backpropagation. This was achieved with the built-in functionality of the PennyLane \texttt{default.qubit} simulator. It is important to emphasize that backpropagation is not feasible on real quantum hardware, where alternative methods, such as parameter-shift or adjoint differentiation~\cite{mitarai2018quantum,jones2020efficient}, must be used.

In addition to the FNN and QML models,
XGBoost regressor is also included for its robustness and strong predictive performance on tabular data. We enabled GPU acceleration using the XGBoost Python library (version 3.0.0). For XGBoost, both the L2 regularization strength and the learning rate were tuned via grid search.

Table~\ref{tab:specs} summarizes the hardware used for model training, optimization procedures, and key hyperparameters.

\begin{table}[htbp]
\resizebox{1.\linewidth}{!}{\begin{tabular}{p{1cm}p{2.5cm}p{2.2cm}p{2.3cm}p{2.2cm}p{1.2cm}p{1.2cm}p{1.5cm}}
\toprule
Model & Device & Module & Optimization & Gradient & Batch Size & Learning Rate \\
\midrule

FNN & GPU & PyTorch & Adam & Backpropagation & 64 & 0.001 \\
XGBoost & GPU & XGBoost & Newton-Raphson & None & None & 0.1 \\

QML & \texttt{default.qubit} state vector simulator on CPU & Pennylane (PyTorch backend) & Adam & Backpropagation & 64 & 0.001 \\
\bottomrule
\end{tabular}}
\caption{Summary of device specifications, optimization protocols, and hyperparameters for each regression model.}
\label{tab:specs}
\end{table}

\color{black}
\subsection{Models' Architecture and Benchmark Metrics}

In our proposed QML architecture, the number of qubits is fixed, with 256 classical features encoded into eight qubits using Amplitude Encoding. While it is theoretically possible to use more qubits, doing so introduces additional complexities, such as an increased number of trainable parameters, the need for input padding during initialization, deeper circuits for state preparation, and potential redundancy in the input representation. Since the simplest configuration—encoding 256 features into eight qubits—yielded satisfactory results, we chose to focus on this setup, leaving the exploration of more complex architectures for future work.

For the alternative models, particularly the FNN and the QML with Angle Encoding, the number of hidden nodes and qubits can be tuned to improve performance. However, to ensure a fair comparison, we selected the number of hidden nodes in the FNN and the number of qubits in the QML with Angle Encoding to match the trainable parameter count of the QML with Amplitude Encoding. Notably, varying these values did not significantly impact performance. A detailed analysis of how performance evolves with the number of hidden nodes in the FNN and qubits in the QML with Angle Encoding is provided in Appendix~\ref{app:fnn} and Appendix~\ref{app:ang} respectively.

For the XGBoost regression model, tree depth was tuned via grid search, with a depth of six yielding the best results. All models were trained for a fixed number of 100 epochs, which for XGBoost corresponds to sequential updates of 100 decision trees. Further performance analysis of XGBoost with varying tree depth and number is presented in Appendix~\ref{app:xgb}. A summary of the model architectures and hyperparameters is provided in Table~\ref{tab:mod_specs}.

\begin{table}[htbp]
    \centering
    \resizebox{1.\linewidth}{!}
    {
    \begin{tabular}{p{1.9cm} p{1.2cm} p{2.cm} p{1.5cm} p{2.3cm} p{2.5cm}}
    \toprule
    Model & Input Layer & I Hidden Layer & Auxiliary Layer & II Hidden Layer & Training Epochs / Trees \\
    \midrule
    FNN               & 256 & 256 LeakyReLU (slope = -0.3) & None & 8 LeakyReLU (slope = -0.3) & 100 Epochs \\
    XGBoost  & 256 & None & None & None & 100 Boosted Trees (depth = 6) \\
    QML Angle Encoding & 256 & 256 LeakyReLU (slope = -0.3) & 8  & 8 Qubits Strongly Entangling PQC & 100 Epochs \\
    QML Amplitude Encoding  & 256 & 256 LeakyReLU (slope = -0.3) & None & 8 Qubits Strongly Entangling PQC & 100 Epochs \\
    \bottomrule
    \end{tabular}
    }
    \caption{Specifications of model architectures, including training epochs or number of trees for each model.}
    \label{tab:mod_specs}
\end{table}

\color{black}
To benchmark the proposed models, we perform four-fold cross-validation, dividing the dataset into four folds, each comprising 75\% training data and 25\% test data. To ensure a fair comparison, all models are evaluated using the same four-fold data splits, and the hyperparameters in Table~\ref{tab:mod_specs} were selected so that the hybrid quantum–classical models and the classical FNN baseline have similar parameter counts. The size of the classical preprocessing layer and the number of trainable variational parameters in the PQC were calibrated to match the effective model capacity across architectures. During training, we record the root mean squared error (RMSE) on the test set for each fold and epoch to monitor the convergence behavior of each model.

Statistical significance testing is essential to ensure that observed performance differences between models reflect meaningful and robust improvements rather than random variation, as emphasized in recent studies such as~\cite{kelly2023modeling, he2024empirical}, and through established methods like the Diebold-Mariano test~\cite{diebold2002comparing}. To assess whether performance differences between models are statistically significant, we conduct a Diebold-Mariano (DM) test, under the assumption of independent forecast errors across observations. For each pair of models (hereafter indicated as $a$ and $b$), we compute the $DM_{a,b}$ statistic across all folds and epochs as follows
\begin{equation}
    DM_{a,b} = \frac{\bar d_{a,b}  }{\sqrt{\mathrm{Var}(\bar d_{a,b})}},
\end{equation}
where $\bar d_{a,b}$ indicates the mean of the differences between the absolute value of the residuals of the model $a$ and $b$, and $\mathrm{Var}(\bar d_{a,b})$ is the estimated variance of the mean difference.
The DM statistic allows us to test whether model $a$ is significantly better than model $b$. A negative value of $DM_{a,b}$ indicates that model $a$ performs better than model $b$, and vice versa for a positive value. Given the large number of test observations (over 400), we compare the DM statistic to the critical values 0.05 from a standard normal distribution to assess the statistical significance of the null hypothesis $DM_{a,b} = 0$, and then conclude that the performance between model $a$ and $b$ is significantly different.

To further strengthen our evaluation, we also perform leave-one-out cross-validation (LOOCV) as an additional robustness check. As with four-fold cross-validation, RMSE is tracked at each epoch for every left-out observation to monitor convergence and model performance.

\subsection{Comparison evaluation with Neural Netwoks}
\label{sec:results_fnn}

\subsubsection{Cross and Leave-One-Out-Cross Validation}

\begin{figure}[htbp]
    \centering

    \begin{subfigure}{0.49\textwidth}
        \centering
        \makebox[\textwidth][l]{\hspace*{-1in}%
            \includegraphics[width=1.25\textwidth]{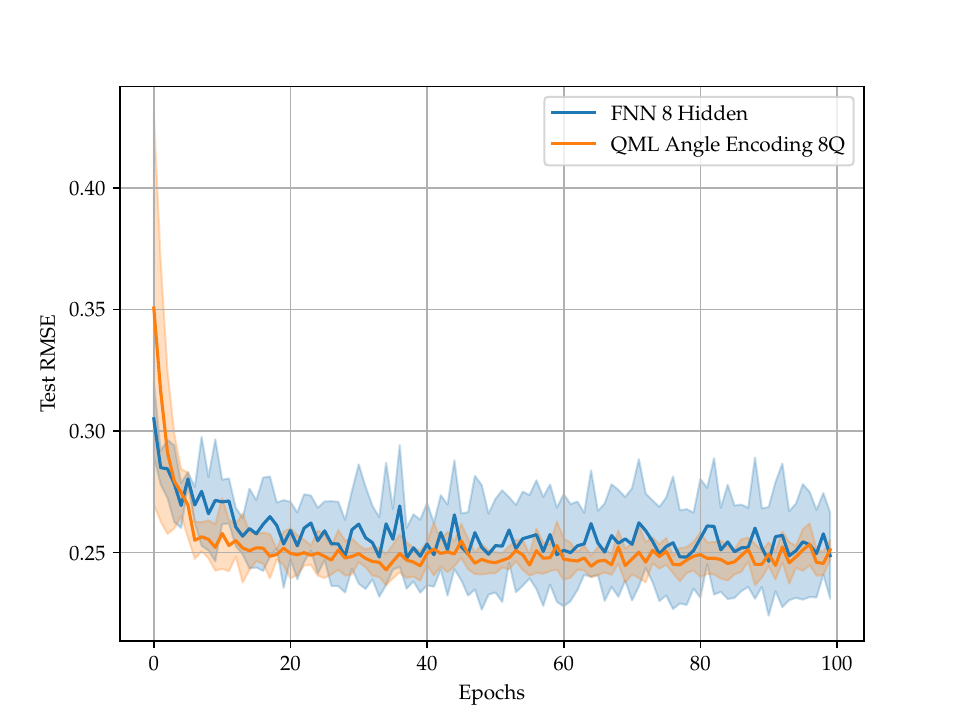}}
        \caption{FNN vs QML (Angle): CV}
        \label{fig:cv_fnn_angle}
    \end{subfigure}
    \hfill
    \begin{subfigure}{0.49\textwidth}
        \centering
        \includegraphics[width=1.25\textwidth]{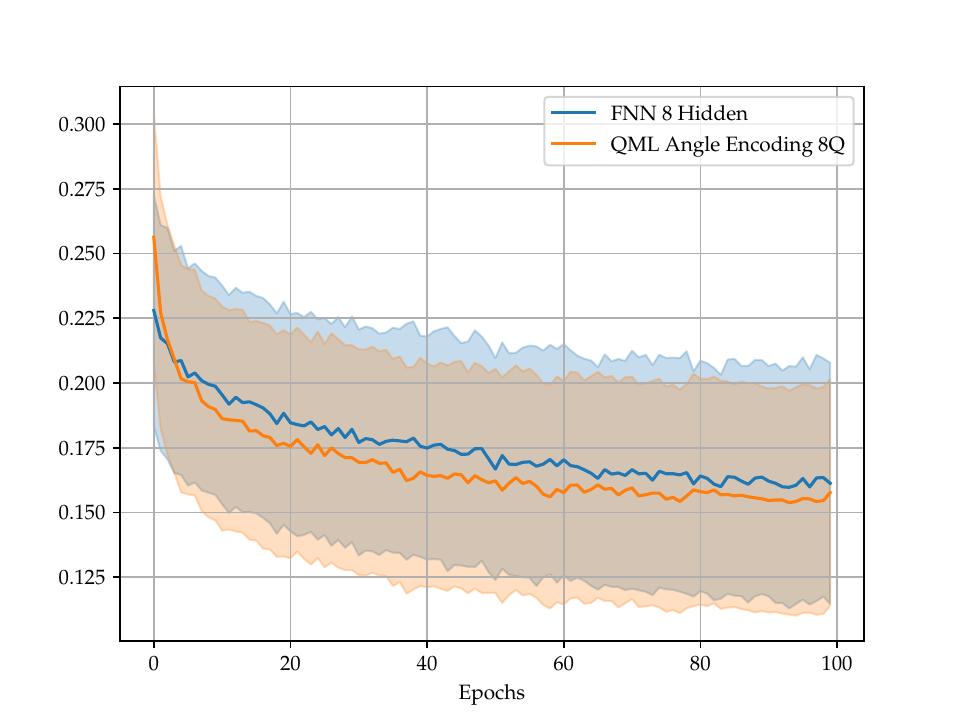}
        \caption{FNN vs QML Angle: LOOCV}
        \label{fig:loocv_fnn_angle}
    \end{subfigure}

    \begin{subfigure}{0.49\textwidth}
        \centering
        \makebox[\textwidth][l]{\hspace*{-1in}%
            \includegraphics[width=1.25\textwidth]{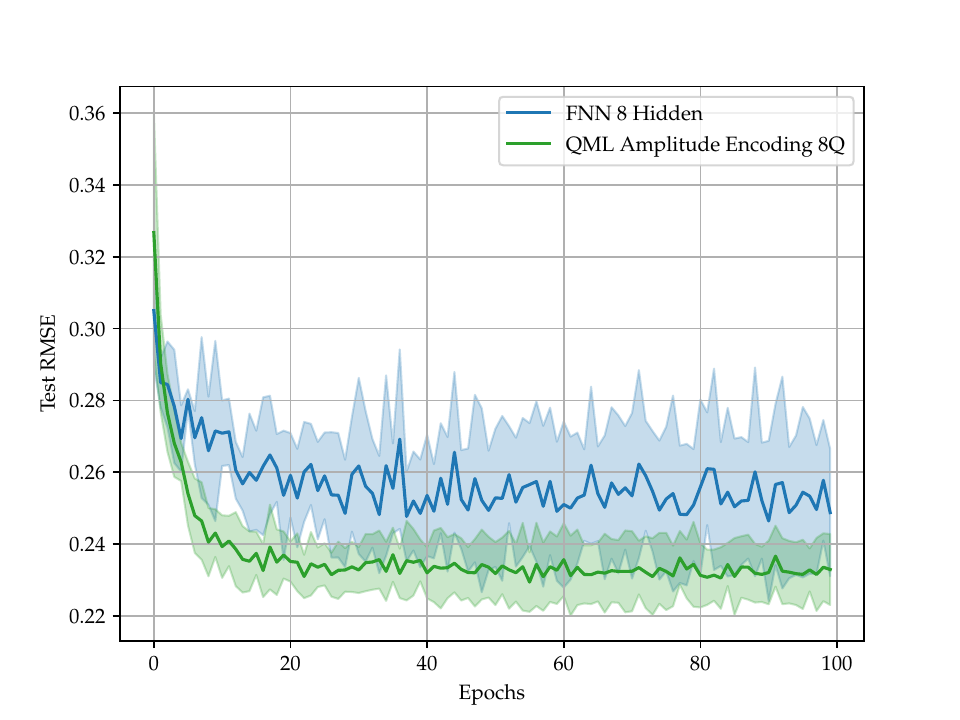}}
        \caption{FNN vs QML Amplitude: CV}
        \label{fig:cv_fnn_amp}
    \end{subfigure}
    \hfill
    \begin{subfigure}{0.49\textwidth}
        \centering
        \includegraphics[width=1.25\textwidth]{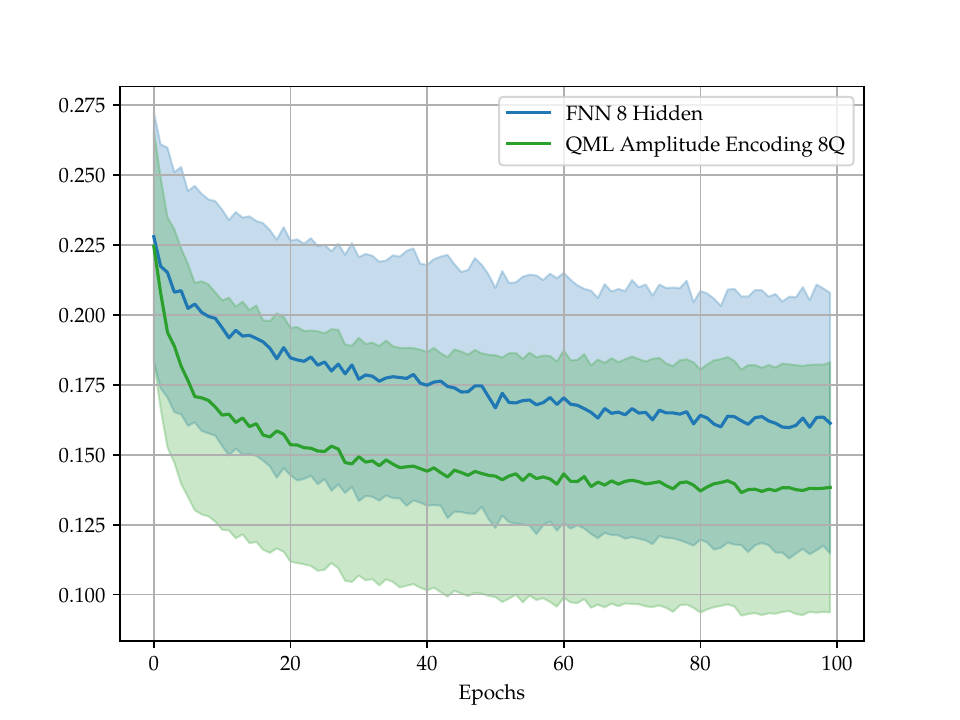}
        \caption{FNN vs QML Amplitude: LOOCV}
        \label{fig:loocv_fnn_amp}
    \end{subfigure}

    \begin{subfigure}{0.49\textwidth}
        \centering
        \makebox[\textwidth][l]{\hspace*{-1in}%
            \includegraphics[width=1.25\textwidth]{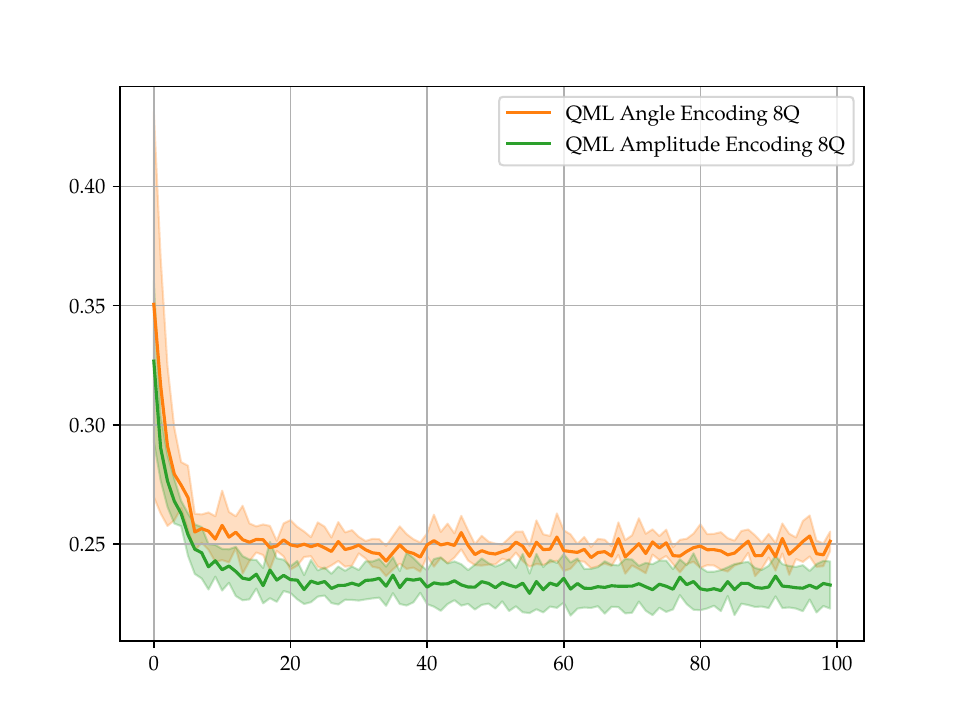}}
        \caption{QML Angle vs QML Amplitude: CV}
        \label{fig:cv_angle_amp}
    \end{subfigure}
    \hfill
    \begin{subfigure}{0.49\textwidth}
        \centering
        \includegraphics[width=1.25\textwidth]{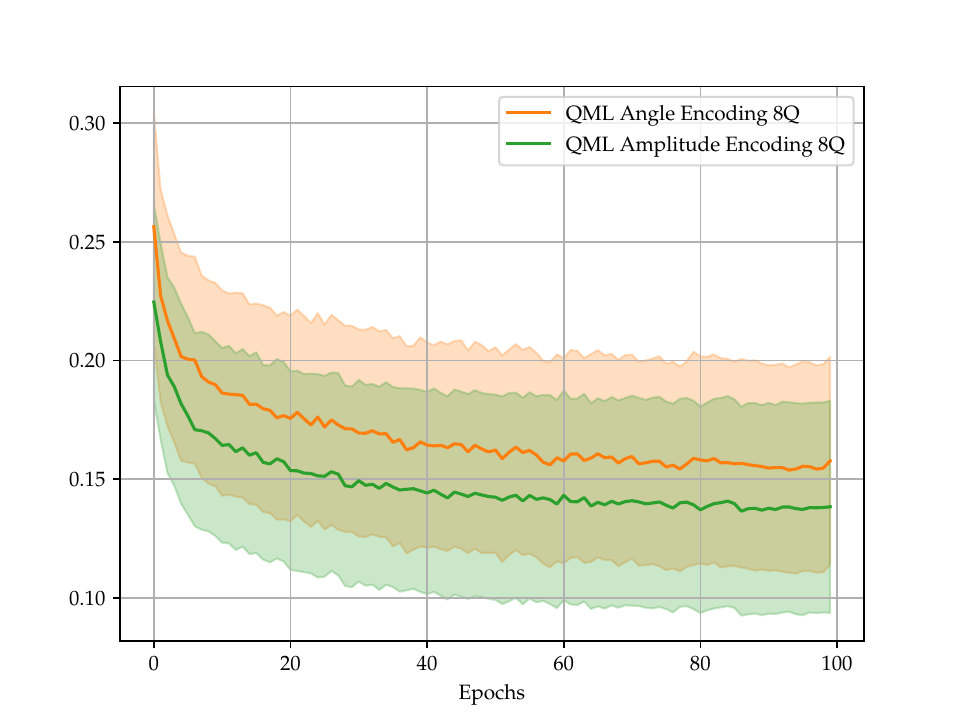}
        \caption{QML Angle vs QML Amplitude: LOOCV}
        \label{fig:loocv_angle_amp}
    \end{subfigure}

    \caption{
        Comparison of four-fold cross-validation (left column, figures (a), (c), and (e)) and leave-one-out cross-validation (right column, figures (b), (d), and (f)) across three pairwise model combinations. 
        Solid lines indicate average test RMSE; shaded regions represent standard deviation. For LOOCV, the shaded region shows a variance scaled by a factor of four for visual clarity.
    }
    \label{fig:all_best}
\end{figure}

Figures~\ref{fig:all_best} show the average RMSE (solid lines) and the corresponding standard deviations (shaded areas) obtained by the FNN, QML with Angle Encoding, and QML with Amplitude Encoding models under four-fold (left column) and leave-one-out cross-validation (right column). Table~\ref{tab:all_best} reports the best average test RMSE achieved during training, the associated average RMSE standard deviation, and the average execution time per epoch.

\begin{table}[htbp]
\centering
\resizebox{1.\linewidth}{!}{
\begin{tabular}{p{2.5cm} p{2.cm} p{2.cm} p{2.2cm} p{1.8cm} p{2.5cm}}
\toprule
Model & \# Parameters & Avg. Time per Epoch (s) & Best Avg. RMSE & At Epoch \# & Avg. STD \\
\midrule
\multirow{2}{*}{FNN} & \multirow{2}{*}{67,857} & \multirow{2}{*}{0.03} & 0.246 (CV) & 90 (CV) & 0.018 (CV) \\
        &           &       & 0.160 (LOOCV) & 93 (LOOCV) & 0.044 (LOOCV) \\
\hline
QML Angle  & \multirow{2}{*}{67,873} & \multirow{2}{*}{0.81} & 0.242 (CV) & 78 (CV) & 0.009 (CV) \\
Encoding   &    &   & 0.153 (LOOCV) & 90 (LOOCV) & 0.044 (LOOCV) \\
\hline
\textbf{QML Amplitude} & \multirow{2}{*}{\textbf{65,825}} & \multirow{2}{*}{\textbf{0.73}} & \textbf{0.228} (CV) & \textbf{55} (CV) & \textbf{0.008} (LOOCV) \\
\textbf{Encoding} &   &       & \textbf{0.136} (LOOCV) & \textbf{86} (LOOCV) & \textbf{0.043} (LOOCV) \\
\bottomrule
\end{tabular}
}
    \caption{Summary of the trainable parameters and key outputs for the selected FNN and QML models, including the best average RMSE, the epoch at which the lowest average RMSE is achieved, the average standard deviation of the RMSE, and the average execution time per epoch. All values are calculated over four cross-validation (CV) folds and leave-one-out (LOOCV), with one hundred epochs.}
    \label{tab:all_best}
\end{table}

The results, summarized in Figure~\ref{fig:all_best} and Tables~\ref{tab:all_best}, provide several key insights. First, the QML models (using Amplitude Encoding and Angle Encoding) outperform the classical FNN model, showcasing superior generalizability. They achieve lower average RMSE in both cross and leave-one-out cross validation, with fewer epochs (see columns three and four of Table~\ref{tab:all_best}). Moreover, their stability is evidenced by the smaller standard deviation observed across the four cross-validation folds and all left-out observations (column five of Table~\ref{tab:all_best}). These findings underscore the effectiveness of replacing a classical layer with a QNN in reducing overfitting, particularly in tasks such as recovery rate prediction.

Second, the QML model with Amplitude Encoding demonstrates the best overall performance. Despite its simplicity and minimal number of trainable parameters, it achieved the lowest average RMSE, requiring the fewest epochs, with the smallest average standard deviation.

\subsubsection{Statistical significance}

\begin{figure}[htbp]
    \centering
        \makebox[\textwidth][l]{\hspace*{-1.3in}\includegraphics[scale=0.43]{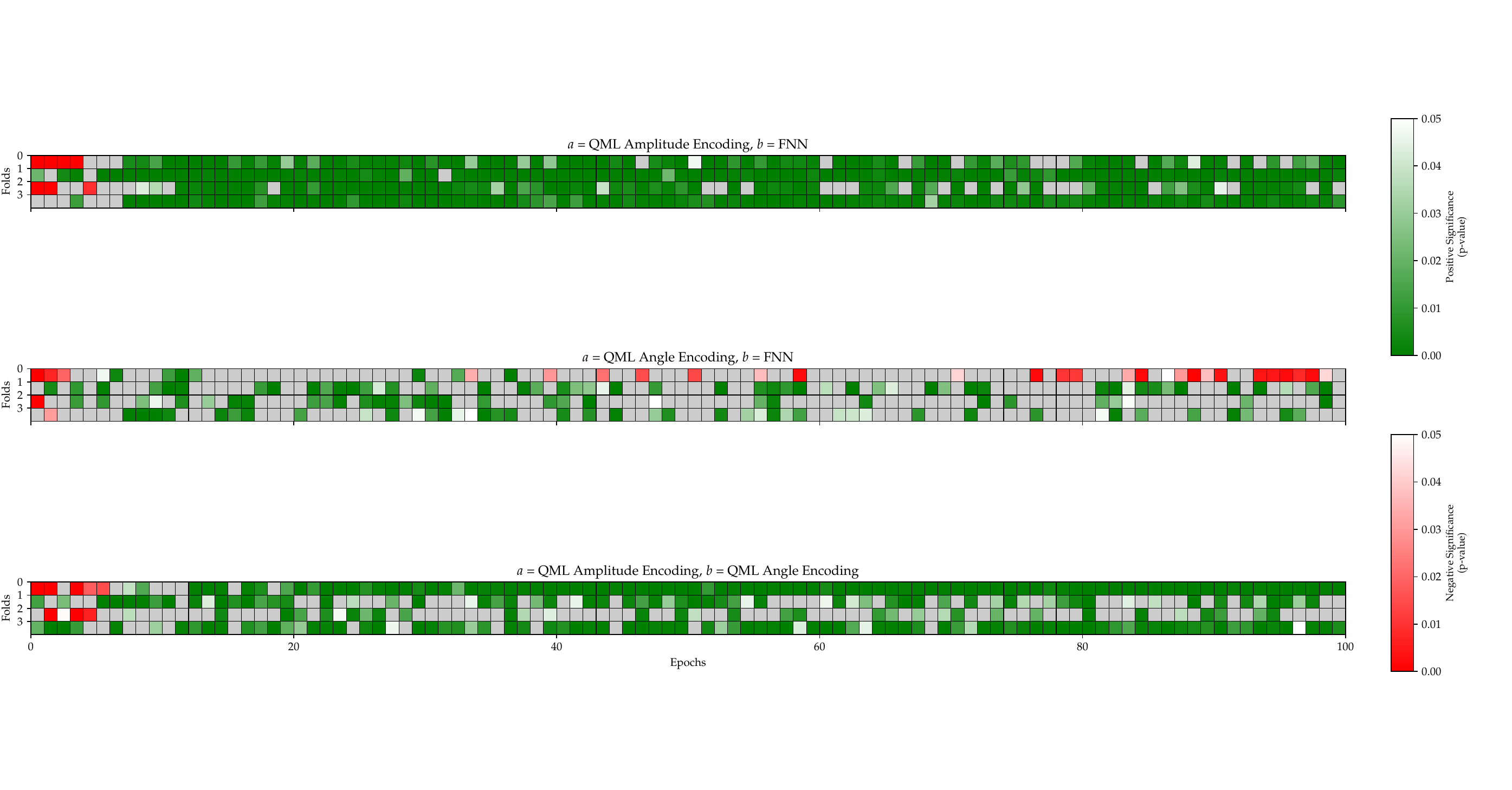}}
        \caption{A summary of test significance across each fold and each epoch. The green (red) boxes indicate that the $DM$ is negative (positive), which means that the model (a) performs better (worse) than model (b). The intensity of the color indicates the p-values: when closer to 0.05, white, which is less significant, and greener (redder) when closer to zero, which indicates higher significance.}
        \label{fig:sign_test}       
\end{figure}

Figure~\ref{fig:sign_test} reports the p-values for the hypothesis test $DM_{a,b} = 0$. The color intensity increases as the p-value approaches zero, indicating a stronger statistical difference between the two models. Boxes are shown in grey when the p-value exceeds the critical threshold of 0.05, suggesting no significant difference between the models. When $DM < 0$ (respectively, $DM > 0$), boxes are colored green (respectively, red), indicating that model $a$ performs significantly better (worse) than model $b$.

Pairwise comparisons of the selected models show that the differences between the QML with Amplitude Encoding and the FNN, and between the QML with Amplitude Encoding and QML with Angle Encoding, are significantly greater than zero across nearly all folds and training epochs (regardless of the starting points in training). This confirms that the QML with Amplitude Encoding model statistically outperforms both the FNN and QML with Angle Encoding models. However, the comparison between FNN and QML with Angle Encoding does not yield consistent results, as the significance tests frequently fail throughout the training process.

In addition to the visual significance map in Figure~\ref{fig:sign_test}, we also summarize the magnitude of the performance differences using fold-averaged RMSE differences and standardized effect sizes based on the results in Table~\ref{tab:all_best}. Across both CV and LOOCV, the QML model with Amplitude Encoding achieves consistently lower RMSE than the FNN and the QML with Angle Encoding, with average improvements ranging from 0.014 to 0.024. Using pooled standard deviations, the corresponding standardized effect sizes fall in the small-to-large range depending on the comparison, indicating that the performance gains are not only statistically significant but also practically meaningful. Since all fold level RMSE differences are negative, the implied confidence intervals for the DM tests do not include zero, which reinforces the robustness of the superiority of the QML model with Amplitude Encoding.

\color{black}
\subsubsection{Discussion}

When comparing the QML models, the Amplitude Encoding model shows more remarkable advantages over Angle Encoding in both stability and accuracy, as detailed in Table~\ref{tab:all_best}, and Figures~\ref{fig:all_best},~\ref{fig:sign_test}. While both models outperform the classical FNN with a comparable number of trainable parameters in terms of average RMSE, the Amplitude Encoding model consistently achieves lower RMSE values and smaller standard deviations across multiple experiments. This superior performance highlights the effectiveness of Amplitude Encoding in capturing and utilizing input information more efficiently. Its enhanced stability and accuracy can be attributed to its ability to represent input data compactly without requiring additional auxiliary layers to reduce input dimensions. By avoiding this added complexity, the Amplitude Encoding model minimizes the risk of overfitting, particularly with limited data. In contrast, the Angle Encoding model relies on an auxiliary classical layer to align the input size with the number of PQC qubits, which can introduce unnecessary complexity and negatively impact performance. Furthermore, the compactness of Amplitude Encoding enables it to encode a larger amount of information into the quantum state using fewer qubits, giving it a significant advantage over Angle Encoding, which scales less efficiently. These attributes make Amplitude Encoding a compelling choice for tasks demanding high accuracy and strong generalization, as reflected in our results.

Finally, we evaluate the computational efficiency of the proposed models. Given the experimental setup, using GPUs for the FNN and a CPU-based quantum simulator for the QML models, direct comparisons of execution times between classical and quantum models are not meaningful. Instead, we focus on the relative performance of the QML models. As shown in the seventh column of Table~\ref{tab:all_best}, the QML model with Amplitude Encoding exhibits slightly better time efficiency than the Angle Encoding model, primarily due to the absence of the auxiliary classical layer in the former.

\subsection{Comparison evaluation with XGBoost}
\label{sec:results_xgb}

\subsubsection{Cross and Leave-One-Out-Cross Validation}

\begin{figure}[htbp]
\begin{subfigure}{.5\textwidth}
    \centering
        \makebox[\textwidth][l]{\hspace*{-1.in}\includegraphics[scale=.6]{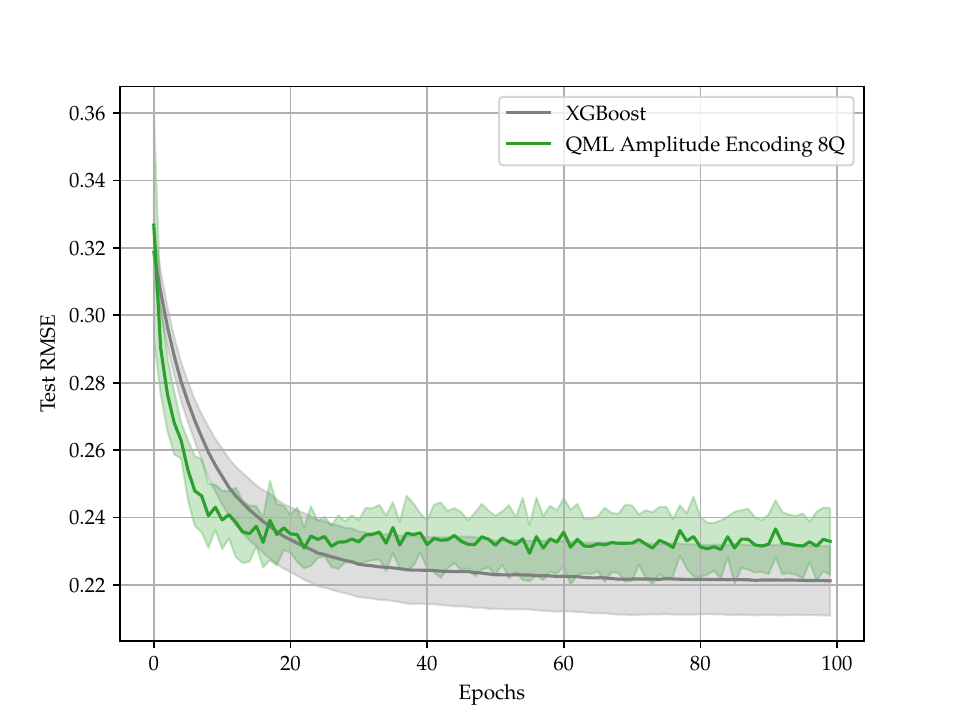}}
        \caption{Four-fold cross validation (CV)}
        \label{fig:cv_xgb_qmlamp}       
\end{subfigure}
\begin{subfigure}{.5\textwidth}
    \centering
    \includegraphics[scale=.6]{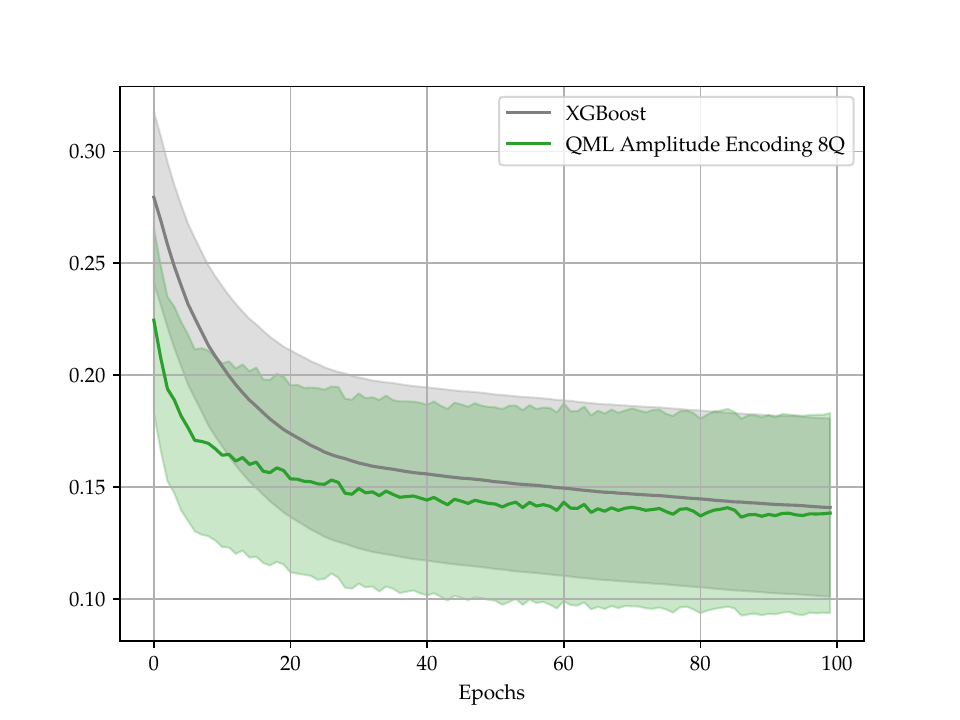}
        \caption{Leave-one-out cross validation (LOOCV)}
        \label{fig:loocv_xgb_qmlAmp}       
\end{subfigure}
\caption{Average test RMSE (solid lines) and standard deviation (shaded area) calculated on a four-fold cross validation (CV left panel) and leave-one-out cross validation (LOOCV right panel) for the XGBoost and QML with Amplitude Encoding. For the sake of visualization, given the high variance of the RMSE calculated on each left-out observation, the shaded area on the right panel indicates a scaled variance by a factor of four.}
\end{figure}

As in the previous section, Figures~\ref{fig:cv_xgb_qmlamp} and~\ref{fig:loocv_xgb_qmlAmp} show the average RMSE (solid lines) and the corresponding standard deviations (shaded areas) obtained by the QML with Amplitude Encoding and XGBoost models under four-fold and leave-one-out cross-validation, respectively. Table~\ref{tab:xgb_qmlamp} reports the best average test RMSE achieved during training, the associated average RMSE standard deviation, and the average execution time per epoch (boosted tree optimization for XGBoost).

\begin{table}[htbp]
\centering
\resizebox{1.\linewidth}{!}{
\begin{tabular}{p{2.5cm} p{2.cm} p{2.cm} p{2.2cm} p{1.9cm} p{2.5cm}}
\toprule
Model & \# Parameters & Avg. Time per Epoch (s) & Best Avg. RMSE & At Epoch/Tree \# & Avg. STD \\
\midrule
\multirow{2}{*}{XGBoost} & \multirow{2}{*}{\textbf{6,400}} & \multirow{2}{*}{0.02} & \textbf{0.222} (CV) & 100 (CV) & 0.010 (CV) \\
        &           &       & 0.141 (LOOCV) & 100 (LOOCV) & \textbf{0.039} (LOOCV) \\
\hline
QML Amplitude & \multirow{2}{*}{65,825} & \multirow{2}{*}{0.73} & 0.228 (CV) & \textbf{55} (CV) & \textbf{0.008} (CV) \\
\textbf{Encoding} &   &       & \textbf{0.136} (LOOCV) & \textbf{86} (LOOCV) & 0.043 (LOOCV) \\
\bottomrule
\end{tabular}
}
    \caption{Summary of the trainable parameters and key outputs for the selected XGBoost and QML with Amplitude Encoding models, including the best average RMSE, the epoch at which the lowest average RMSE is achieved, the average standard deviation of the RMSE, and the average execution time per epoch. All values are calculated over four cross-validation (CV) folds and leave-one-out (LOOCV), with one hundred epochs (trees).}
    \label{tab:xgb_qmlamp}
\end{table}

In this case, as shown in Figure~\ref{fig:cv_xgb_qmlamp} and Table~\ref{tab:xgb_qmlamp}, the XGBoost regression model achieves a lower RMSE value on the four-fold cross-validation setting. However, as we can see from column four of Table~\ref{tab:xgb_qmlamp}, to achieve the convergence XGBoost model needs multiple trees, potentially compromising the performance of the model (a deeper discussion about the scalability and the convergence of XGBoost is provided in Appendix~\ref{app:xgb}). Additionally, as illustrated in Figure~\ref{fig:loocv_xgb_qmlAmp} and column three of Table~\ref{tab:xgb_qmlamp} in the leave-one-out validation, after one hundred boosted trees, the average RMSE is still higher than the one achieved by our proposed QML model.

\subsubsection{Statistical significance}

\begin{figure}[htbp]
    \centering
        \makebox[\textwidth][l]{\hspace*{-1.9in}\includegraphics[scale=0.55]{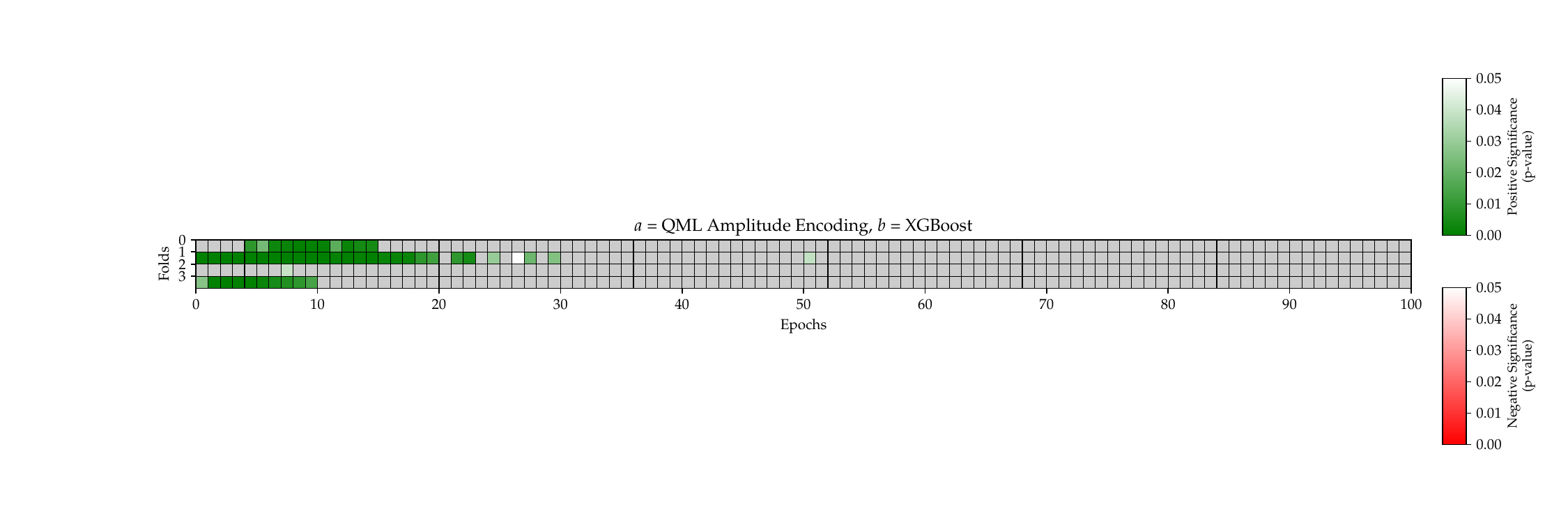}}
        \caption{A summary of test significance across each fold and each epoch. The green (red) boxes indicate that the $DM$ is negative (positive), which means that the model (a) performs better (worse) than model (b). The intensity of the color indicates the p-values: when closer to 0.05, white, which is less significant, and greener (redder) when closer to zero, which indicates higher significance.}
        \label{fig:sign_test_xgb}       
\end{figure}

As in the previous section, Figure~\ref{fig:sign_test_xgb} highlights the instances where the QML model with Amplitude Encoding significantly outperforms the XGBoost model. Although XGBoost achieved lower RMSE values, its higher variance prevents it from being statistically significantly better than the QML model with Amplitude Encoding across most epochs and folds.
To quantify the magnitude of these differences, we also compare the fold-averaged RMSE values reported in Table~\ref{tab:xgb_qmlamp}. The average RMSE gap between the two models is small in magnitude, ranging from $+0.006$ under CV to $-0.005$ under LOOCV, and the corresponding standardized effect sizes fall in the small-to-medium range depending on the evaluation setting. Because the RMSE differences fluctuate around zero across folds, the implied confidence intervals for the Diebold--Mariano statistics necessarily include zero, which confirms that neither model consistently dominates the other in a statistically robust sense. These observations align with Figure~\ref{fig:sign_test_xgb}, where significance appears sporadically rather than persistently across epochs and folds.

\subsubsection{Discussion}

The comparative analysis of XGBoost and QML with Amplitude Encoding models reveals a nuanced performance landscape. As shown in Figure~\ref{fig:cv_xgb_qmlamp} and Table~\ref{tab:xgb_qmlamp}, in standard cross-validation, XGBoost achieved a lower average RMSE compared to QML, though this came at the cost of higher variance, which increases during training. Conversely, under leave-one-out cross-validation, the QML model slightly outperforms XGBoost in terms of RMSE.

When examining performance across training epochs for QML and the sequential addition of trees in XGBoost, no statistically significant advantage consistently emerged for XGBoost over the proposed QML model, as pictured in Figure~\ref{fig:sign_test_xgb}. Notably, in the early stages of training, QML demonstrated a statistically significant advantage over XGBoost, suggesting faster initial learning.

Column three of Table~\ref{tab:xgb_qmlamp} shows that, while XGBoost exhibited faster execution times, this comparison is not meaningful: XGBoost was executed on GPU hardware, whereas QML relied on classical quantum simulation. Moreover, as illustrated in column~\ref{tab:xgb_qmlamp}, although XGBoost typically requires fewer tunable parameters per tree, its convergence depends on a large number of boosted trees. This iterative process increases the overall model complexity and the effective number of parameters (further discussion about the scalability of XGBoost is reported in Appendix~\ref{app:xgb}).

Taken together, these findings suggest that, despite current hardware limitations, QML remains a viable and competitive approach for recovery rate prediction. With further in-depth research, fine-tuning, and architectural adaptation, its complexity could potentially be reduced and its performance enhanced.

\subsection{Further analysis about performance differences across Neural Network models}

In this section, we highlight key architectural and training-related features of the proposed QML model with Amplitude Encoding that contribute to its competitive performance compared to both classical baselines and QML models using Angle Encoding.

First, Amplitude Encoding allows the QNN to represent all input features simultaneously in a compact quantum state. Unlike Angle Encoding—which encodes each feature individually via rotation gates and often requires dimensionality reduction to fit within qubit number constraints—Amplitude Encoding enables the representation of $2^n$ features using only $n$ qubits. This efficiency eliminates the need for auxiliary preprocessing steps and preserves more information, which is particularly beneficial for high-dimensional, small-sample tasks.

Second, the QML with Amplitude Encoding introduces a Parameterized Quantum Circuit (PQC) as a unitary transformation within the neural architecture. This unitary layer contributes to better generalization, reducing the risk of overfitting~\cite{bartlett2017spectrally, li2019orthogonal}. Unlike classical Feedforward Neural Network (FNN), which often requires more complex architectures and additional constraints to prevent overfitting, the QNN benefits from a more compact yet expressive representation.

Third, the number of trainable parameters in the amplitude-encoded QNN scales logarithmically with the number of input features. This is in contrast to classical FNNs and angle-encoded QNN, where parameter count typically scales linearly with input size. The resulting architecture is not only more compact but also exhibits more stable training dynamics. 

Together, these factors explain the superior performance of the QML model with Amplitude Encoding in both predictive accuracy and robustness.

\section{Conclusion and Discussion}
\label{sec:concl}

In this work, we proposed and evaluated a QML model with Amplitude Encoding for recovery rate prediction, comparing its performance with a QML model using Angle Encoding, classical Feedforward Neural Network (FNN) and XGBoost. 

Our experimental results show that the QML model with Amplitude Encoding consistently outperforms the QML model with Angle Encoding and classical FNN in terms of accuracy, stability, and generalization. Its lower RMSE, statistically significant reduction in prediction error, and smaller standard deviation across multiple training runs and cross-validation settings highlight its robustness and reduced tendency to overfit, making it a promising candidate for complex prediction tasks.
Although the QML model with Angle Encoding shows improvements over the classical FNN, this is not as significant as the one achieved by the QML with Amplitude encoding. The performance of the QML with Angle Encoding is hindered by the added complexity of the auxiliary classical layer, which increases the risk of overfitting. In contrast, the simplicity of the Amplitude Encoding model, with fewer layers and parameters, allows it to achieve better results with a more stable training process.


When compared to XGBoost, the proposed QML model with Amplitude Encoding does not demonstrate clearly superior performance. In fact, it yields a higher average RMSE in cross-validation settings, albeit with a smaller standard deviation. However, due to the higher variance observed in XGBoost, its performance improvement is not statistically significant. Moreover, in the leave-one-out cross-validation, the proposed QML model outperforms XGBoost. These findings suggest that further tuning of the QML model’s parameters and architecture could potentially enhance its performance and make it more competitive.

A direct comparison across the models clarifies the trade-offs between predictive accuracy and computational efficiency. As summarized in Table \ref{tab:all_best}, the QML model with Amplitude Encoding achieves the lowest average RMSE and smallest standard deviation among the neural-network–based architectures, while the FNN and XGBoost trains substantially faster in simulation (around 0.03/0.02 s per epoch versus approximately 0.73 s for the hybrid QML model). The Angle-Encoding variant lies in between: it is slower than the FNN due to the additional auxiliary layer and exhibits higher variance than the Amplitude-Encoding model. These results highlight the central trade-off of hybrid quantum models in the NISQ era—improved predictive regularization and stability at the cost of greater computational overhead.

From a computational standpoint, the QML model with Amplitude Encoding is also slightly more time-efficient in simulation than the Angle-Encoding variant, as expected from its simpler structure. However, a direct runtime comparison between the quantum and classical models is meaningless in simulation, since the classical architectures are trained on a GPU, whereas the quantum model runs on a CPU-based quantum simulation, as explicitly underlined in~\autoref{tab:specs}. A fair assessment of time efficiency can only be made once the QML model is executed on real quantum hardware. Moreover, while the Amplitude-Encoding model shows better simulated performance, its implementation on actual quantum devices remains challenging due to the deep circuits required, which are more prone to noise and decoherence (see Section \ref{sec:limit}, and supplementary materials for further details).


\subsection{Limitation}
\label{sec:limit}

While this study demonstrates the potential of QML models in recovery rate modeling, several limitations remain, particularly in relation to hardware implementation, noise handling, feature encoding strategies, and overall Machine Learning architecture.

First, all QML models in this paper are developed and tested using quantum simulators rather than real quantum hardware. Simulation allows for noiseless training and evaluation, which does not fully reflect the limitations of existing quantum devices. In practice, current quantum processors suffer from limited qubit counts, gate fidelity issues, and short coherence times. As a result, implementing QNNs on real hardware remains a significant challenge. For example, Angle Encoding strategies typically require more qubits but produce shallower circuits. At the same time, Amplitude Encoding is more qubit-efficient but results in deeper and more complex circuits that are more sensitive to noise~\cite{mottonen2004transformation, shende2005synthesis, weigold2021expanding}. 

Second, our current implementation does not explicitly model or mitigate noise effects such as decoherence and gate errors. These sources of quantum noise can significantly deteriorate model performance when deploying on real quantum hardware. Techniques such as noise-aware training, error mitigation, and quantum error correction are potential directions for future work, especially as quantum hardware matures~\cite{wang2024artificial, qin2022overview}. However, a preliminary evaluation of the proposed models on real quantum hardware and noisy classical simulators is provided in Appendix~\ref{app:QPU}. Although these results are still in the early stages and do not yet support strong conclusions, they indicate that the proposed models may offer sufficient noise resilience and retain promising performance even under realistic noise conditions. 

Third, the encoding of high-dimensional features into quantum circuits presents practical challenges. Amplitude Encoding requires the feature vector to be of length $2^n$, necessitating the use of padding or zero-filling techniques when the number of features does not match a power of two~\cite{schuld2020circuit}. While padding allows compatibility, it may introduce redundancy or distort input distributions. Alternatively, Angle Encoding avoids this restriction but becomes impractical when the number of features is very large, as it requires one qubit per feature. In such cases, dimensionality reduction techniques or hybrid classical-quantum approaches are needed to reduce the input space before quantum processing.

In addition to the input formatting challenges, the number of qubits required for Amplitude Encoding still grows logarithmically with the number of input features. While this scaling is significantly more efficient than linear methods such as Angle Encoding, it can still become a bottleneck in practice, given the limited number of qubits currently available on Noisy Intermediate-Scale Quantum (NISQ) devices. For example, encoding 1,024 features would require at least 10 qubits, which does not account for additional qubits needed for variational processing or error mitigation. As a result, even moderate feature dimensions may exceed the capabilities of current hardware. This constraint highlights the need for preprocessing strategies such as dimensionality reduction or feature grouping prior to quantum encoding, or the development of more hardware-efficient encoding methods.

Moreover, the high dimensionality of the input is initially processed by a classical master layer, which may reduce the quantum speedup benefits by introducing computational bottlenecks. This reflects a broader trade-off in hybrid quantum-classical architectures. While the current design represents one viable approach, it is not necessarily optimal.
Like classical neural networks, our QML model inherits challenges such as reduced interpretability and increased computational cost during training and inference on high-dimensional datasets~\cite{hastie2009elements, goodfellow2016deep}.

Finally, an additional limitation relates to the economic significance of the performance gains achieved by the QML model. Although the hybrid quantum model yields a 7--8\% reduction in RMSE relative to a strong classical FNN baseline, the economic value of such improvements in recovery-rate prediction remains context dependent. Recovery modeling is inherently noisy, and even small improvements can translate into non-trivial differences in expected loss, stress-testing outcomes, or economic-capital calculations. However, in the current technological environment, quantum models incur substantially higher computational costs, particularly when using state preparation through Amplitude Encoding or when relying on noisy simulations or early-stage hardware. As a result, the present study should be interpreted as a proof of concept demonstrating methodological feasibility rather than a cost-efficient alternative to classical machine-learning systems. Assessing the full economic cost–benefit trade-off of QML for credit-risk modeling will require future advances in both hardware scalability and quantum runtime efficiency.

Overall, these limitations highlight both the promise and the current barriers to deploying QML in real-world financial applications. Overcoming these challenges will be essential as quantum technologies advance toward practical deployment.

\subsection{Future Works}

Building on the current findings, several directions for future research emerge, both from a methodological and an applied financial perspective.

First, future work can address the limitations discussed above by exploring more robust and efficient quantum model implementations. In particular, alternative encoding strategies such as basis encoding, hybrid schemes~\cite{ranga2024quantum}, or data re-uploading~\cite{perez2021one} could be evaluated for their effectiveness and scalability. When using Amplitude Encoding, padding strategies for non-power-of-two feature dimensions could be systematically tested to assess their impact on model accuracy and stability. Similarly, further exploration of dimensionality reduction techniques would be valuable when using Angle Encoding with high-dimensional inputs. From a hardware perspective, the deployment of QNNs on real quantum devices will require advances in circuit optimization, efficient gradient calculation (e.g., parameter-shift rules~\cite{mitarai2018quantum}, adjoint differentiation~\cite{jones2020efficient}), and noise mitigation strategies to ensure reliable training and inference. A practical roadmap toward real-device deployment will also depend on access to low-noise hardware with sufficient logical qubits (on the order of 8--12), support for deeper state-preparation circuits, and the use of approximate or truncated amplitude-encoding schemes combined with error-mitigation and compilation techniques. Although runtime comparisons between classical models and hybrid quantum architectures are in principle possible, current quantum runtimes are dominated by device-level overhead (compilation, queuing, noise mitigation) rather than intrinsic algorithmic complexity; a meaningful runtime assessment is therefore left for future work, once more scalable and stable hardware becomes available.

Second, the economic implications of improved recovery rate prediction warrant further investigation. A natural extension would be to evaluate whether the QML-based recovery forecasts can inform profitable trading strategies. For instance, one could simulate a strategy that selectively purchases defaulted bonds with high predicted recovery rates and avoids or shorts those with low expected recoveries. However, the feasibility and profitability of such strategies depend on market efficiency, liquidity, and transaction costs. It is possible that bond prices already reflect recovery expectations due to market participants' access to similar information. Therefore, future work should explore whether the QML model provides incremental economic value beyond existing credit risk models and whether its predictions offer statistically and economically significant alpha.

Third, an important direction for future research is to evaluate the model's adaptability to dynamic financial environments. In our current study, the rarity of default events and limited sample size make it difficult to assess temporal generalization directly. However, as more recovery data become available or as higher-frequency credit risk datasets (e.g., CDS spreads, market-based distress indicators) are integrated, future studies could investigate whether QML models maintain predictive performance across different market regimes or macroeconomic cycles. One possible approach is to incorporate time-varying covariates or macro factors into the model architecture, or to test model transferability across temporal subsamples. Additionally, evaluating the robustness of QML under concept drift or structural shifts in default behavior would help assess their practical viability in real-world, evolving financial systems.

Fourth, it would be natural to extend the present framework beyond standalone recovery-rate prediction to more integrated credit-risk applications. Examples include joint modeling of probability of default (PD), exposure at default (EAD), and loss given default (LGD) within a unified quantum-enhanced architecture, or embedding variational quantum kernels into structural or reduced-form credit-risk models. Such extensions would require richer datasets and additional methodological development, but they offer a promising avenue for leveraging quantum machine learning in portfolio-level risk management and regulatory capital assessment.

Together, these directions highlight the dual potential of QML for both technical innovation and practical impact in credit risk modeling.

\backmatter

\color{black}
\section*{Acknowledgements}

The authors gratefully acknowledge the support of Whitespace under grant number A-0003504-17-00 and the Quantum Engineering Program (QEP) under grant number A-8000339-01-00. We would also like to thank the Amazon Braket team for their help throughout the project in implementing on the real QPU Rigetti Ankaa-3.

\section*{Declarations}

The codes used in this study are available from the corresponding author upon reasonable request.

The data used to train the models presented in this work were obtained through the NRF Research Project UP5 at the National University of Singapore and are subject to a nondisclosure agreement.


\bibliography{quantum_recovery}

@inproceedings{hardt2016train,
  title={Train faster, generalize better: Stability of stochastic gradient descent},
  author={Hardt, Moritz and Recht, Ben and Singer, Yoram},
  booktitle={International conference on machine learning},
  pages={1225--1234},
  year={2016},
  organization={PMLR}
}

@article{jankowitsch2014determinants,
  title={The determinants of recovery rates in the US corporate bond market},
  author={Jankowitsch, Rainer and Nagler, Florian and Subrahmanyam, Marti G},
  journal={Journal of Financial Economics},
  volume={114},
  number={1},
  pages={155--177},
  year={2014},
  publisher={Elsevier}
}

@article{he2024empirical,
  title={Empirical Asset Pricing with Probability Forecasts},
  author={He, Songrun and Lv, Linying and Zhou, Guofu},
  journal={Available at SSRN},
  year={2024}
}

@article{kelly2023modeling,
  title={Modeling corporate bond returns},
  author={Kelly, Bryan and Palhares, Diogo and Pruitt, Seth},
  journal={The Journal of Finance},
  volume={78},
  number={4},
  pages={1967--2008},
  year={2023},
  publisher={Wiley Online Library}
}

@article{gambetti2022meta,
  title={Meta-Learning Approaches for Recovery Rate Prediction},
  author={Gambetti, Sara and Roccazzella, Lorenzo and Vrins, Frédéric},
  journal={Journal of Machine Learning in Finance},
  volume={5},
  number={2},
  pages={107--128},
  year={2022},
  publisher={World Scientific}
}

@article{altman2005link,
  title={The Link between Default and Recovery Rates: Theory, Empirical Evidence, and Implications},
  author={Altman, Edward I and Resti, Andrea and Sironi, Andrea},
  journal={The Journal of Business},
  volume={78},
  number={6},
  pages={2203--2227},
  year={2005},
  publisher={University of Chicago Press}
}

@article{schuld2020circuit,
  title={Circuit-centric quantum classifiers},
  author={Schuld, Maria and Bocharov, Alex and Svore, Krysta M and Wiebe, Nathan},
  journal={Physical Review A},
  volume={101},
  number={3},
  pages={032308},
  year={2020},
  publisher={APS}
}

@article{perez2021one,
  title={One qubit as a universal approximant},
  author={Pérez-Salinas, Adrián and Cervera-Lierta, Alba and Gil-Fuster, Enrique and Latorre, José I},
  journal={Physical Review A},
  volume={104},
  number={1},
  pages={012405},
  year={2021},
  publisher={APS}
}

@article{li2019orthogonal,
  title={Orthogonal deep neural networks},
  author={Li, Shuai and Xu, Wei and Liu, Shiji and Lin, Xin and Zhang, Hongwei},
  journal={IEEE Transactions on Pattern Analysis and Machine Intelligence},
  volume={43},
  number={4},
  pages={1352--1368},
  year={2019},
  publisher={IEEE}
}

@inproceedings{shende2005synthesis,
  title={Synthesis of quantum logic circuits},
  author={Shende, Vivek V and Bullock, Stephen S and Markov, Igor L},
  booktitle={Proceedings of the 2005 Asia and South Pacific Design Automation Conference},
  pages={272--275},
  year={2005}
}

@inproceedings{weigold2021expanding,
  title={Expanding data encoding patterns for quantum algorithms},
  author={Weigold, Manuela and Barzen, Johanna and Leymann, Frank and Salm, Marie},
  booktitle={2021 IEEE 18th International Conference on Software Architecture Companion (ICSA-C)},
  pages={95--101},
  year={2021},
  organization={IEEE}
}

@article{wang2024artificial,
  title={Artificial Intelligence for Quantum Error Correction: A Comprehensive Review},
  author={Wang, Zihao and Tang, Hao},
  journal={arXiv preprint arXiv:2412.20380},
  year={2024}
}

@article{qin2022overview,
  title={An overview of quantum error mitigation formulas},
  author={Qin, Dayue and Xu, Xiaosi and Li, Ying},
  journal={Chinese Physics B},
  volume={31},
  number={9},
  pages={090306},
  year={2022},
  publisher={IOP Publishing}
}

@article{nassirtoussi2015text,
  title={Text mining of news-headlines for FOREX market prediction: A Multi-layer Dimension Reduction Algorithm with semantics and sentiment},
  author={Nassirtoussi, Arman Khadjeh and Aghabozorgi, Saeed and Wah, Teh Ying and Ngo, David Chek Ling},
  journal={Expert Systems with Applications},
  volume={42},
  number={1},
  pages={306--324},
  year={2015},
  publisher={Elsevier}
}

@book{hastie2009elements,
  title={The elements of statistical learning: data mining, inference, and prediction},
  author={Hastie, Trevor and Tibshirani, Robert and Friedman, Jerome H and Friedman, Jerome H},
  volume={2},
  year={2009},
  publisher={Springer}
}

@article{nassirtoussi2014text,
  title={Text mining for market prediction: A systematic review},
  author={Nassirtoussi, Arman Khadjeh and Aghabozorgi, Saeed and Wah, Teh Ying and Ngo, David Chek Ling},
  journal={Expert Systems with Applications},
  volume={41},
  number={16},
  pages={7653--7670},
  year={2014},
  publisher={Elsevier}
}

@article{giesecke2011corporate,
  title={Corporate bond default risk: A 150-year perspective},
  author={Giesecke, Kay and Longstaff, Francis A and Schaefer, Stephen and Strebulaev, Ilya},
  journal={Journal of financial Economics},
  volume={102},
  number={2},
  pages={233--250},
  year={2011},
  publisher={Elsevier}
}

@article{sim2019expressibility,
  title={Expressibility and entangling capability of parameterized quantum circuits for hybrid quantum-classical algorithms},
  author={Sim, Sukin and Johnson, Peter D and Aspuru-Guzik, Alán},
  journal={Advanced Quantum Technologies},
  volume={2},
  number={12},
  pages={1900070},
  year={2019},
  publisher={Wiley}
}

@article{benedetti2019parameterized,
  title={Parameterized quantum circuits as machine learning models},
  author={Benedetti, Marcello and Lloyd, Seth and Sack, Stefan and Fiorentini, Mattia},
  journal={Quantum Science and Technology},
  volume={4},
  number={4},
  pages={043001},
  year={2019},
  publisher={IOP Publishing}
}

@article{ranga2024quantum,
  title={Quantum Machine Learning: Exploring the Role of Data Encoding Techniques, Challenges, and Future Directions},
  author={Ranga, Deepak and Rana, Aryan and Prajapat, Sunil and Kumar, Pankaj and Kumar, Kranti and Vasilakos, Athanasios V},
  journal={Mathematics},
  volume={12},
  number={21},
  pages={3318},
  year={2024},
  publisher={MDPI}
}

@article{mottonen2004transformation,
  title={Transformation of quantum states using uniformly controlled rotations},
  author={Mottonen, Mikko and Vartiainen, Juha J and Bergholm, Ville and Salomaa, Martti M},
  journal={arXiv preprint quant-ph/0407010},
  year={2004}
}

@article{rath2024quantum,
  title={Quantum data encoding: a comparative analysis of classical-to-quantum mapping techniques and their impact on machine learning accuracy},
  author={Rath, Minati and Date, Hema},
  journal={EPJ Quantum Technology},
  volume={11},
  number={1},
  pages={72},
  year={2024},
  publisher={Springer Berlin Heidelberg}
}

@article{gong2024quantum,
  title={Quantum convolutional neural network based on variational quantum circuits},
  author={Gong, Li-Hua and Pei, Jun-Jie and Zhang, Tian-Feng and Zhou, Nan-Run},
  journal={Optics Communications},
  volume={550},
  pages={129993},
  year={2024},
  publisher={Elsevier}
}

@article{montanaro2015quantum,
  title={Quantum speedup of Monte Carlo methods},
  author={Montanaro, Ashley},
  journal={Proceedings of the Royal Society A: Mathematical, Physical and Engineering Sciences},
  volume={471},
  number={2181},
  pages={20150301},
  year={2015},
  publisher={The Royal Society}
}

@article{plekhanov2022variational,
  title={Variational quantum amplitude estimation},
  author={Plekhanov, Kirill and Rosenkranz, Matthias and Fiorentini, Mattia and Lubasch, Michael},
  journal={Quantum},
  volume={6},
  pages={670},
  year={2022},
  publisher={Verein zur F{\"o}rderung des Open Access Publizierens in den Quantenwissenschaften}
}

@book{schuld2018supervised,
  title={Supervised learning with quantum computers},
  author={Schuld, Maria},
  year={2018},
  publisher={Springer}
}

@article{mitarai2018quantum,
  title={Quantum circuit learning},
  author={Mitarai, Kosuke and Negoro, Makoto and Kitagawa, Masahiro and Fujii, Keisuke},
  journal={Physical Review A},
  volume={98},
  number={3},
  pages={032309},
  year={2018},
  publisher={APS}
}

@article{jones2020efficient,
  title={Efficient calculation of gradients in classical simulations of variational quantum algorithms},
  author={Jones, Tyson and Gacon, Julien},
  journal={arXiv preprint arXiv:2009.02823},
  year={2020}
}

@article{kingma2014adam,
  title={Adam: A method for stochastic optimization},
  author={Kingma, Diederik P},
  journal={arXiv preprint arXiv:1412.6980},
  year={2014}
}

@misc{pennylane_plugins,
  author = {Pennylane},
  title = {Pennylane devices and ecosystem},
  howpublished = {\url{https://pennylane.ai/plugins}}
}

@misc{pennylane_defaultQ,
  author = {Pennylane},
  title = {Pennylane default.qubit},
  howpublished = {\url{https://docs.pennylane.ai/en/stable/code/api/pennylane.devices.default_qubit.html}}
}

@article{alcazar2020classical,
  title={Classical versus quantum models in machine learning: insights from a finance application},
  author={Alcazar, Javier and Leyton-Ortega, Vicente and Perdomo-Ortiz, Alejandro},
  journal={Machine Learning: Science and Technology},
  volume={1},
  number={3},
  pages={035003},
  year={2020},
  publisher={IOP Publishing}
}

@article{liu2022quantum,
  title={A quantum artificial neural network for stock closing price prediction},
  author={Liu, Ge and Ma, Wenping},
  journal={Information Sciences},
  volume={598},
  pages={75--85},
  year={2022},
  publisher={Elsevier}
}

@article{emmanoulopoulos2022quantum,
  title={Quantum machine learning in finance: Time series forecasting},
  author={Emmanoulopoulos, Dimitrios and Dimoska, Sofija},
  journal={arXiv preprint arXiv:2202.00599},
  year={2022}
}

@inproceedings{rivera2022time,
  title={Time series forecasting with quantum machine learning architectures},
  author={Rivera-Ruiz, Mayra Alejandra and Mendez-Vazquez, Andres and L{\'o}pez-Romero, Jos{\'e} Mauricio},
  booktitle={Mexican International Conference on Artificial Intelligence},
  pages={66--82},
  year={2022},
  organization={Springer}
}

@article{bartlett2017spectrally,
  title={Spectrally-normalized margin bounds for neural networks},
  author={Bartlett, Peter L and Foster, Dylan J and Telgarsky, Matus J},
  journal={Advances in neural information processing systems},
  volume={30},
  year={2017}
}

@article{bansal2018can,
  title={Can we gain more from orthogonality regularizations in training deep networks?},
  author={Bansal, Nitin and Chen, Xiaohan and Wang, Zhangyang},
  journal={Advances in Neural Information Processing Systems},
  volume={31},
  year={2018}
}

@inproceedings{cisse2017parseval,
  title={Parseval networks: Improving robustness to adversarial examples},
  author={Cisse, Moustapha and Bojanowski, Piotr and Grave, Edouard and Dauphin, Yann and Usunier, Nicolas},
  booktitle={International conference on machine learning},
  pages={854--863},
  year={2017},
  organization={PMLR}
}

@article{bartlett1996valid,
  title={For valid generalization the size of the weights is more important than the size of the network},
  author={Bartlett, Peter},
  journal={Advances in neural information processing systems},
  volume={9},
  year={1996}
}

@article{mottonen2004quantum,
  title={Quantum circuits for general multiqubit gates},
  author={M{\"o}tt{\"o}nen, Mikko and Vartiainen, Juha J and Bergholm, Ville and Salomaa, Martti M},
  journal={Physical review letters},
  volume={93},
  number={13},
  pages={130502},
  year={2004},
  publisher={APS}
}

@inproceedings{wang2020orthogonal,
  title={Orthogonal convolutional neural networks},
  author={Wang, Jiayun and Chen, Yubei and Chakraborty, Rudrasis and Yu, Stella X},
  booktitle={Proceedings of the IEEE/CVF conference on computer vision and pattern recognition},
  pages={11505--11515},
  year={2020}
}

@article{kyriienko2022unsupervised,
  title={Unsupervised quantum machine learning for fraud detection},
  author={Kyriienko, Oleksandr and Magnusson, Einar B},
  journal={arXiv preprint arXiv:2208.01203},
  year={2022}
}

@inproceedings{tekkali2023smart,
  title={Smart Payment Fraud Detection using QML--A Major Challenge},
  author={Tekkali, Chandana Gouri and Natarajan, Karthika},
  booktitle={2023 Third International Conference on Artificial Intelligence and Smart Energy (ICAIS)},
  pages={523--526},
  year={2023},
  organization={IEEE}
}

@article{schetakis2024quantum,
  title={Quantum machine learning for credit scoring},
  author={Schetakis, Nikolaos and Aghamalyan, Davit and Boguslavsky, Michael and Rees, Agnieszka and Rakotomalala, Marc and Griffin, Paul Robert},
  journal={Mathematics},
  volume={12},
  number={9},
  pages={1391},
  year={2024},
  publisher={MDPI}
}

@article{brassard2002quantum,
  title={Quantum amplitude amplification and estimation},
  author={Brassard, Gilles and Hoyer, Peter and Mosca, Michele and Tapp, Alain},
  journal={Contemporary Mathematics},
  volume={305},
  pages={53--74},
  year={2002},
  publisher={Providence, RI; American Mathematical Society; 1999}
}

@article{zoufal2019quantum,
  title={Quantum generative adversarial networks for learning and loading random distributions},
  author={Zoufal, Christa and Lucchi, Aur{\'e}lien and Woerner, Stefan},
  journal={npj Quantum Information},
  volume={5},
  number={1},
  pages={103},
  year={2019},
  publisher={Nature Publishing Group UK London}
}

@article{schuld2019evaluating,
  title={Evaluating analytic gradients on quantum hardware},
  author={Schuld, Maria and Bergholm, Ville and Gogolin, Christian and Izaac, Josh and Killoran, Nathan},
  journal={Physical Review A},
  volume={99},
  number={3},
  pages={032331},
  year={2019},
  publisher={APS}
}

@article{Nazemi2018improving,
  author    = {Abdolreza Nazemi and Konstantin Heidenreich and Frank Fabozzi},
  title     = {Improving Corporate Bond Recovery Rate Prediction Using Multi-Factor Support Vector Regressions},
  journal   = {European Journal of Operational Research},
  year      = {2018},
  volume    = {271},
  pages     = {664--675},
  doi       = {10.1016/j.ejor.2018.05.045}
}

@article{nazemi2018macroeconomic,
  author = {Nazemi, Amir and Fabozzi, Frank J.},
  title = {Macroeconomic factors in predicting corporate bond recovery rates: A LASSO approach},
  journal = {Journal of Fixed Income},
  volume = {27},
  number = {4},
  pages = {55--74},
  year = {2018}
}

@article{nazemi2022machine,
  author = {Nazemi, Amir and Fabozzi, Frank J.},
  title = {Comparing machine learning techniques and statistical models for bond recovery rate forecasting},
  journal = {Journal of Financial Data Science},
  volume = {4},
  number = {2},
  pages = {1--20},
  year = {2022}
}

@article{mashhadi2021parallel,
  title={Parallel orthogonal deep neural network},
  author={Mashhadi, Peyman Sheikholharam and Nowaczyk, S{\l}awomir and Pashami, Sepideh},
  journal={Neural Networks},
  volume={140},
  pages={167--183},
  year={2021},
  publisher={Elsevier}
}

@inproceedings{henaff2016recurrent,
  title={Recurrent orthogonal networks and long-memory tasks},
  author={Henaff, Mikael and Szlam, Arthur and LeCun, Yann},
  booktitle={International Conference on Machine Learning},
  pages={2034--2042},
  year={2016},
  organization={PMLR}
}

@article{le2015simple,
  title={A simple way to initialize recurrent networks of rectified linear units},
  author={Le, Quoc V and Jaitly, Navdeep and Hinton, Geoffrey E},
  journal={arXiv preprint arXiv:1504.00941},
  year={2015}
}

@misc{Rigetti,
  author = {Rigetti},
  title = {Rigetti Systems Ankaa-3 Quantum Processor},
  howpublished = {\url{https://qcs.rigetti.com/qpus}}
}

@article{diebold2002comparing,
  title={Comparing predictive accuracy},
  author={Diebold, Francis X and Mariano, Robert S},
  journal={Journal of Business \& economic statistics},
  volume={20},
  number={1},
  pages={134--144},
  year={2002},
  publisher={Taylor \& Francis}
}

@article{das2009implied,
  title={Implied recovery},
  author={Das, Sanjiv R and Hanouna, Paul},
  journal={Journal of Economic Dynamics and Control},
  volume={33},
  number={11},
  pages={1837--1857},
  year={2009},
  publisher={Elsevier}
}

@techreport{BCBS2023,
  author       = {{Basel Committee on Banking Supervision}},
  title        = {Basel Framework - Calculation of RWA for Credit Risk - IRB Approach: Risk Components},
  year         = {2023},
  month        = {December},
  institution  = {Bank for International Settlements},
  address      = {Basel, Switzerland},
  url          = {https://www.bis.org/basel_framework/chapter/CRE/32.htm?inforce=20230101&published=20200327}
}

@article{cantor2007moody,
  title={Moody’s Ultimate Recovery Database: Special Report},
  author={Cantor, R and Emery, K and Keisman, D and Ou, S},
  journal={Moody’s Investor Service},
  year={2007}
}

@article{Pykthin2003,
  author    = {Michael Pykthin},
  title     = {Unexpected Recovery Risk},
  journal   = {Risk},
  year      = {2003},
  volume    = {16},
  pages     = {74--78}
}

@article{Andersen2004,
  author    = {Leif Andersen and Jakob Sidenius},
  title     = {Extensions to the Gaussian Copula: Random Recovery and Random Factor Loadings},
  journal   = {Journal of Credit Risk},
  year      = {2004},
  volume    = {1},
  pages     = {29--70},
  doi       = {10.21314/JCR.2004.001}
}

@article{Berd2005,
  author    = {Arthur Berd},
  title     = {Recovery Swaps},
  journal   = {Journal of Credit Risk},
  year      = {2005},
  volume    = {1},
  pages     = {61--70},
  doi       = {10.21314/JCR.2005.008}
}

@incollection{Gambetti2018,
  author    = {Paolo Gambetti and Geneviève Gauthier and Frédéric Vrins},
  title     = {Stochastic Recovery Rate: Impact of Pricing Measure’s Choice and Financial Consequences on Single-Name Products},
  booktitle = {New Methods in Fixed Income Analysis},
  editor    = {Mehdi Mili and Reyes Samaniego Medina and Filippo Di Pietro},
  publisher = {Springer},
  address   = {Cham},
  year      = {2018},
  pages     = {181--203},
  doi       = {10.1007/978-3-319-96032-3_9}
}

@techreport{Moodys2011,
  author       = {{Moody’s}},
  title        = {Sovereign Default and Recovery Rates, 1983-2010},
  year         = {2011},
  type         = {Special Comment},
  institution  = {Moody’s Investors Service}
}

@article{altman1996defaulted,
  author = {Altman, Edward I. and Kishore, Vellore M.},
  title = {Defaulted debt recovery rates of public utilities and industrial firms},
  journal = {Journal of Applied Corporate Finance},
  volume = {9},
  number = {4},
  pages = {88--98},
  year = {1996}
}

@article{acharya2007creditor,
  author = {Acharya, Viral V. and Bharath, Sreedhar T. and Srinivasan, Arun},
  title = {Does industry-wide distress affect defaulted firms? Evidence from the credit market},
  journal = {Journal of Financial Economics},
  volume = {85},
  number = {2},
  pages = {452--479},
  year = {2007}
}

@article{altman2014mixture,
  author = {Altman, Edward I. and Kalotay, Eric A.},
  title = {A mixture model for estimating recovery rates of defaulted debt},
  journal = {Journal of Fixed Income},
  volume = {23},
  number = {2},
  pages = {40--50},
  year = {2014}
}

@article{jankowitsch2014recovery,
  author = {Jankowitsch, Rainer and Lando, David and S{\o}rensen, Lars},
  title = {Recovery rates and credit spreads: A study of defaulted bonds},
  journal = {Journal of Financial Economics},
  volume = {112},
  number = {2},
  pages = {221--243},
  year = {2014}
}

@article{qi2011comparison,
  title={Comparison of modeling methods for loss given default},
  author={Qi, Min and Zhao, Xinlei},
  journal={Journal of Banking \& Finance},
  volume={35},
  number={11},
  pages={2842--2855},
  year={2011},
  publisher={Elsevier}
}

@article{liu2024machine,
  title={Machine learning prediction of loss given default in government-sponsored enterprise residential mortgages},
  author={Liu, Zilong and Liang, Hongyan},
  journal={Journal of Risk Model Validation},
  year={2024}
}

@article{yao2017enhancing,
  title={Enhancing two-stage modelling methodology for loss given default with support vector machines},
  author={Yao, Xiao and Crook, Jonathan and Andreeva, Galina},
  journal={European Journal of Operational Research},
  volume={263},
  number={2},
  pages={679--689},
  year={2017},
  publisher={Elsevier}
}

@article{kalotay2017intertemporal,
  title={Intertemporal forecasts of defaulted bond recoveries and portfolio losses},
  author={Kalotay, Egon A and Altman, Edward I},
  journal={Review of Finance},
  volume={21},
  number={1},
  pages={433--463},
  year={2017},
  publisher={Oxford University Press}
}

@article{schuld2014quest,
  title={The quest for a quantum neural network},
  author={Schuld, Maria and Sinayskiy, Ilya and Petruccione, Francesco},
  journal={Quantum Information Processing},
  volume={13},
  pages={2567--2586},
  year={2014},
  publisher={Springer}
}

@article{biamonte2017quantum,
  title={Quantum machine learning},
  author={Biamonte, Jacob and Wittek, Peter and Pancotti, Nicola and Rebentrost, Patrick and Wiebe, Nathan and Lloyd, Seth},
  journal={Nature},
  volume={549},
  number={7671},
  pages={195--202},
  year={2017},
  publisher={Nature Publishing Group UK London}
}

@article{abbas2021power,
  title={The power of quantum neural networks},
  author={Abbas, Amira and Sutter, David and Zoufal, Christa and Lucchi, Aur{\'e}lien and Figalli, Alessio and Woerner, Stefan},
  journal={Nature Computational Science},
  volume={1},
  number={6},
  pages={403--409},
  year={2021},
  publisher={Nature Publishing Group US New York}
}

@inproceedings{lezcano2019cheap,
  title={Cheap orthogonal constraints in neural networks: A simple parametrization of the orthogonal and unitary group},
  author={Lezcano-Casado, Mario and Mart{\i}nez-Rubio, David},
  booktitle={International Conference on Machine Learning},
  pages={3794--3803},
  year={2019},
  organization={PMLR}
}

@article{tibshirani1996regression,
  title={Regression shrinkage and selection via the lasso},
  author={Tibshirani, Robert},
  journal={Journal of the Royal Statistical Society Series B: Statistical Methodology},
  volume={58},
  number={1},
  pages={267--288},
  year={1996},
  publisher={Oxford University Press}
}

@inproceedings{chen2016xgboost,
  title={Xgboost: A scalable tree boosting system},
  author={Chen, Tianqi and Guestrin, Carlos},
  booktitle={Proceedings of the 22nd acm sigkdd international conference on knowledge discovery and data mining},
  pages={785--794},
  year={2016}
}

@article{havlicek2019supervised,
  title={Supervised learning with quantum-enhanced feature spaces},
  author={Havlíček, Vojtěch and Córcoles, Antonio D. and Temme, Kristan and Harrow, Aram W. and Kandala, Abhinav and Chow, Jerry M. and Gambetta, Jay M.},
  journal={Nature},
  volume={567},
  pages={209--212},
  year={2019},
  doi={10.1038/s41586-019-0980-2}
}

@article{schuld2019quantum,
  title={Quantum machine learning in feature Hilbert spaces},
  author={Schuld, Maria and Killoran, Nathan},
  journal={Physical review letters},
  volume={122},
  number={4},
  pages={040504},
  year={2019},
  publisher={APS}
}

@article{calabrese2014predicting,
  author = {Calabrese, Raffaella and Zenga, Mariangela},
  title = {Predicting bank loan recovery rates with ordered logit models},
  journal = {Journal of the Operational Research Society},
  year = {2014},
  volume = {65},
  number = {3},
  pages = {393--407}
}

@article{dermine2006bank,
  title={Bank loan losses-given-default: A case study},
  author={Dermine, Jean and De Carvalho, C Neto},
  journal={Journal of Banking \& Finance},
  volume={30},
  number={4},
  pages={1219--1243},
  year={2006},
  publisher={Elsevier}
}

@article{duan2012public,
  title={A public good approach to credit ratings--from concept to reality},
  author={Duan, Jin-Chuan and Van Laere, Elisabeth},
  journal={Journal of Banking \& Finance},
  volume={36},
  number={12},
  pages={3239--3247},
  year={2012},
  publisher={Elsevier}
}

@article{duan2012multiperiod,
  title={Multiperiod corporate default prediction—A forward intensity approach},
  author={Duan, Jin-Chuan and Sun, Jie and Wang, Tao},
  journal={Journal of Econometrics},
  volume={170},
  number={1},
  pages={191--209},
  year={2012},
  publisher={Elsevier}
}

@article{duan2014predicting,
  author = {Duan, Jin-Chuan and Hwang, Min},
  title = {Predicting recovery rate at the time of corporate default},
  journal = {Working Paper},
  year = {2014},
  note = {Available at \url{https://d.nuscri.org/profiles/duanjc/index_files/files/ConditionalRecoveryRate_Nov-13-2018.pdf}}
}

@article{bellotti2012loss,
  title={Loss given default models incorporating macroeconomic variables for credit cards},
  author={Bellotti, Tony and Crook, Jonathan},
  journal={International Journal of Forecasting},
  volume={28},
  number={1},
  pages={171--182},
  year={2012},
  publisher={Elsevier}
}

@article{duffie2009frailty,
  title={Frailty correlated default},
  author={Duffie, Darrell and Eckner, Andreas and Horel, Guillaume and Saita, Leandro},
  journal={The Journal of Finance},
  volume={64},
  number={5},
  pages={2089--2123},
  year={2009},
  publisher={Wiley Online Library}
}

@article{friewald2012illiquidity,
  title={Illiquidity or credit deterioration: A study of liquidity in the US corporate bond market during financial crises},
  author={Friewald, Nils and Jankowitsch, Rainer and Subrahmanyam, Marti G},
  journal={Journal of financial economics},
  volume={105},
  number={1},
  pages={18--36},
  year={2012},
  publisher={Elsevier}
}

@techreport{hamilton2001default,
  author = {Hamilton, David T.},
  title = {Default and Recovery Rates of Corporate Bond Issuers: 2000},
  institution = {Moody's Investors Service},
  year = {2001},
  month = {February},
  note = {Special Comment},
  url = {https://ssrn.com/abstract=277999}
}

@article{frye2000loss,
  author = {Frye, Jon},
  title = {Loss given default as a function of the default rate},
  journal = {Federal Reserve Bank of Chicago Working Paper},
  year = {2000},
  note = {Available at \url{https://www.chicagofed.org/-/media/others/people/research-resources/frye-jon/frye-lgd-as-a-function-of-the-default-rate-091013-pdf.pdf}}
}

@article{franks2004recovery,
  author = {Franks, Julian R. and Torous, Walter N.},
  title = {A study of inefficient going concerns in bankruptcy},
  journal = {Journal of Finance},
  year = {2004},
  volume = {53},
  number = {3},
  pages = {949--969}
}

@article{varma2005determinants,
  author = {Varma, Prem and Cantor, Richard},
  title = {Determinants of Recovery Rates on Defaulted Bonds and Loans for North American Corporate Issuers: 1983-2003},
  journal = {Moody's Investors Service},
  year = {2005},
  note = {Special Comment}
}

@article{yuan2006model,
  author = {Yuan, Ming and Lin, Yi},
  title = {Model selection and estimation in regression with grouped variables},
  journal = {Journal of the Royal Statistical Society: Series B (Statistical Methodology)},
  year = {2006},
  volume = {68},
  number = {1},
  pages = {49--67}
}

@article{jacobs2011modeling,
  author = {Jacobs, Michael and Karagozoglu, Ahmet K.},
  title = {Modeling ultimate loss given default on corporate debt},
  journal = {Journal of Fixed Income},
  year = {2011},
  volume = {21},
  number = {2},
  pages = {67--82}
}

@article{donovan2015accounting,
  author = {Donovan, John and Frankel, Richard and Martin, Xiumin},
  title = {Accounting conservatism and creditor recovery rates},
  journal = {Journal of Accounting and Economics},
  year = {2015},
  volume = {60},
  number = {2-3},
  pages = {202--220}
}

@article{mora2015determinants,
  author = {Mora, Nada},
  title = {The determinants of bank loan recovery rates in good times and bad},
  journal = {Journal of Financial Intermediation},
  year = {2015},
  volume = {24},
  number = {4},
  pages = {522--544}
}

@article{bertsimas2016best,
  author = {Bertsimas, Dimitris and King, Andrew and Mazumder, Rahul},
  title = {Best subset selection via a modern optimization lens},
  journal = {The Annals of Statistics},
  year = {2016},
  volume = {44},
  number = {2},
  pages = {813--852}
}

@book{goodfellow2016deep,
  author = {Goodfellow, Ian and Bengio, Yoshua and Courville, Aaron},
  title = {Deep Learning},
  publisher = {MIT Press},
  year = {2016},
  address = {Cambridge, MA}
}

@article{kim2016determinants,
  author = {Kim, Jin-Chul and Kung, Hsiu-Ching},
  title = {Determinants of recovery rates on defaulted bonds: Evidence from the Korean bond market},
  journal = {Emerging Markets Finance and Trade},
  year = {2016},
  volume = {52},
  number = {5},
  pages = {1107--1120}
}

@article{lundberg2017unified,
  author = {Lundberg, Scott M. and Lee, Su-In},
  title = {A unified approach to interpreting model predictions},
  journal = {Advances in Neural Information Processing Systems},
  year = {2017},
  volume = {30},
  pages = {4765--4774}
}

@article{araci2019finbert,
  author = {Araci, Dogu},
  title = {FinBERT: Financial sentiment analysis with pre-trained language models},
  journal = {arXiv preprint arXiv:1908.10063},
  year = {2019}
}

@article{gambetti2019modeling,
  author = {Gambetti, Luca and Golinelli, Roberto and Parigi, Giuseppe},
  title = {Modeling recovery rates on defaulted bonds: A mixture model approach},
  journal = {Journal of Empirical Finance},
  year = {2019},
  volume = {53},
  pages = {1--17}
}


\begin{appendices}

\color{black}
\section{Scalability of alternative regression models}

In this appendix, we examine how the alternative models—FNN, QML with Angle Encoding, and XGBoost—respond to increasing architectural complexity. Specifically, we analyze the impact of increasing the number of hidden nodes in the FNN, the number of qubits in the QML model, and both the depth and number of trees in XGBoost.

\subsection{The Classical FNNs and overfitting}
\label{app:fnn}

\begin{figure}[htbp]
    \centering
        \includegraphics[scale=0.6]{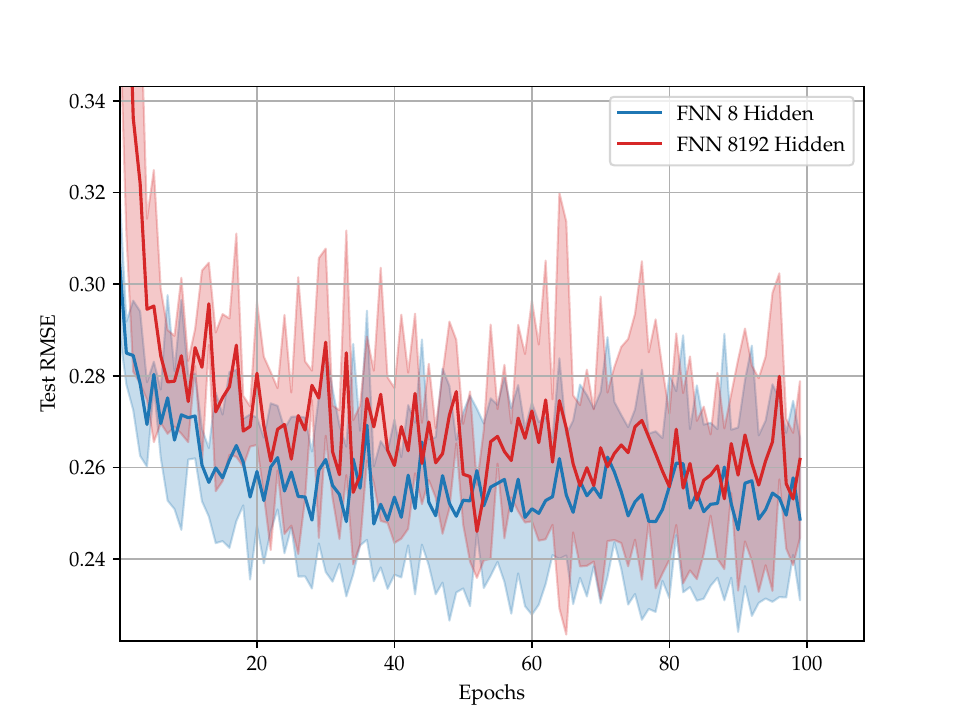}
        \caption{The average test RMSE and STD calculated over ten different experiments of the two extreme FNN configuration; specifically the one with 8 hidden nodes and the one with 8192.}
        \label{fig:CML_16_8192}       
    \end{figure}

In this section, we evaluate the performance of various FNN models, by increasing the number of hidden nodes. Figure~\ref{fig:CML_16_8192} illustrates the average RMSE and standard deviation on test data across ten experiments for two extreme configurations: one with 8 hidden nodes and another with 8192 hidden nodes. Additionally, the table presents the lowest RMSE and corresponding standard deviation observed over ten experiments for each configuration.

   \begin{table}[htbp]
    \centering
    \begin{tabular}{lllll}
    \toprule
    Model & $n$ Hidden    & $\#$ Parameters   & Best Avg. RMSE \#  & Avg. STD \\
    \midrule
    \multirow{6}{*}{QML Angle Encoding} & 8           & 67,857 & 0.246 & 0.018 \\
                                        & 16          & 69,921 & 0.244 & 0.017 \\
                                        & 128         & 98,817 & 0.244 & 0.023 \\
                                        & 512         & 197,889& 0.246 & 0.021 \\
                                        & 2048        & 594,177& 0.246 & 0.024 \\
                                        & 8192        &2,179,329& 0.250 & 0.031 \\
    \hline
    {\bf QML Amplitude Encoding}&None & {\bf 65,825} & {\bf 0.228} & {\bf 0.008} \\
    \bottomrule
    \end{tabular}
    \caption{The total number of trainable parameters and some relevant outputs related to the different FNN configurations, specifically the best average RMSE, and the best STD calculated over the four cross-validation folds with one hundred epochs.}
    \label{tab:cml_comp}
    \end{table}    

From Figure~\ref{fig:CML_16_8192} and Table~\ref{tab:cml_comp}, it is evident that increasing the number of hidden nodes does not improve the FNN's prediction accuracy or stability, as indicated by both the RMSE and standard deviation. In fact, further increasing the number of hidden nodes worsens both prediction performance and stability. As discussed in Section~\ref{sec:data}, the recovery rate prediction problem is particularly susceptible to overfitting, and adding more complexity to the model is counterproductive.


\subsection{Performance increasing the number of qubits in the QML model with Angle Encoding}
\label{app:ang}

\begin{figure}[htbp]
\begin{subfigure}{.5\textwidth}
    \centering
    \makebox[\textwidth][l]{\hspace*{-1.in}\includegraphics[scale=.6]
        {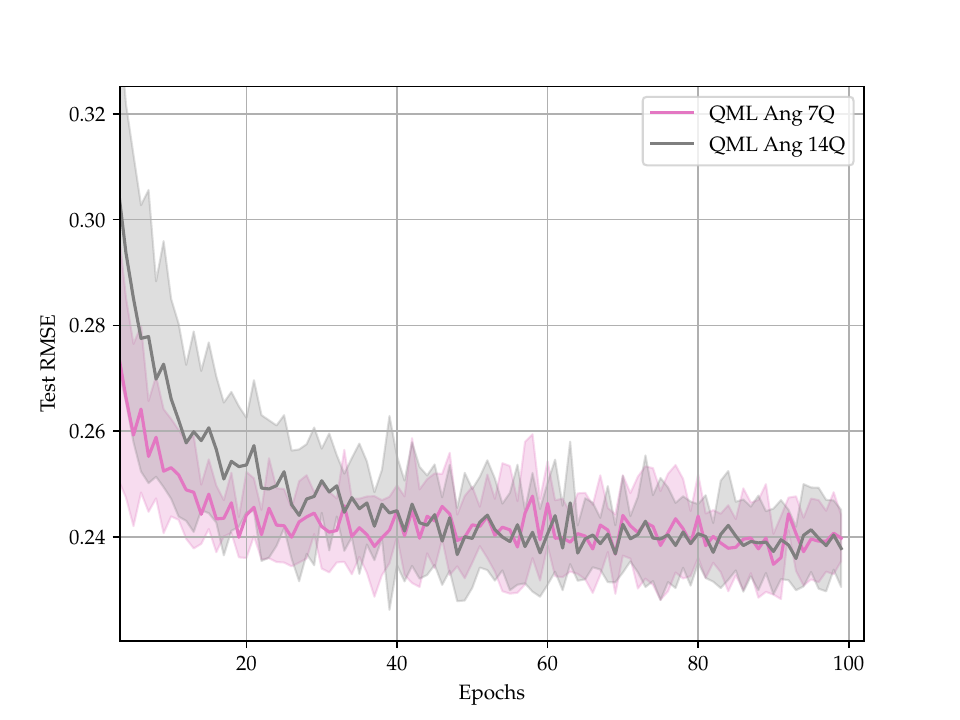}}
        \caption{Test RMSE}
        \label{fig:QML_ang_7_14}
\end{subfigure}
\begin{subfigure}{0.5\textwidth}
    \centering
     \includegraphics[scale=.6]{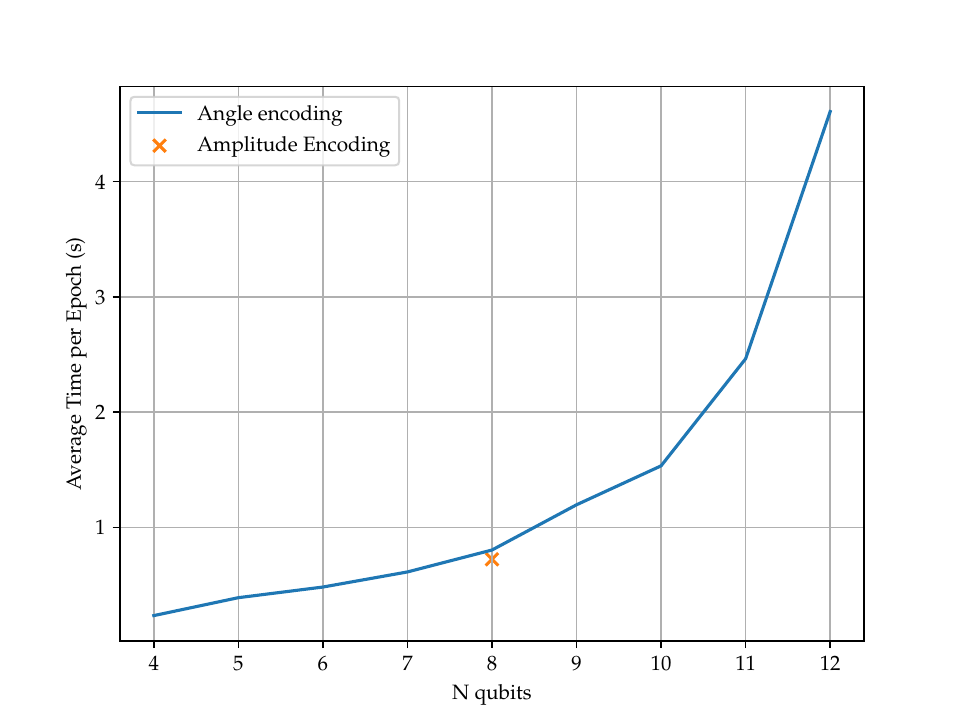}
    \caption{Average Time per epoch}
    \label{fig:time_ang_amp}
\end{subfigure}
\caption{(Left panel) The average test RMSE and STD calculated over four corss-validation folds of the QML with Angle Encoding with seven and fourteen qubits in the PQC. (Right Panel) Average execution time per epoch in the QML with Angle Encoding (blue line) and Amplitude Encoding (orange marker).}
\end{figure}

In the architecture proposed in Section~\ref{sec:meth}, the number of qubits can be adjusted to identify the optimal configuration for the QML model using Angle Encoding. Figure~\ref{fig:QML_ang_7_14} presents the average RMSE and standard deviation on the test data for two configurations: one with six qubits and another with fourteen qubits in the PQC. As in the previous section, Table~\ref{tab:qml_ang_comp} summarizes the best RMSE and corresponding standard deviation values observed across ten experiments for different qubit configurations.

\begin{table}[htbp]
    \centering
    \begin{tabular}{lllll}
    \toprule
    Model   & $n$ Qubits    & $\#$ Parameters   & Best Avg. RMSE & Avg. STD \\
    \midrule
    \multirow{6}{*}{QML Angle Encoding} & 6         & 67,353 & 0.241  & 0.012 \\
                                        & 7         & 67,613 & 0.241  & 0.009 \\
                                        & 8         & 67,873 & 0.242  & 0.009 \\
                                        & 10        & 68,393 & 0.242  & 0.011 \\
                                        & 12        & 68,913 & 0.242  & 0.010 \\
                                        & 14        & 69,433 & 0.242  & 0.013 \\
    \hline
    {\bf QML Amplitude Encoding} & 8 & {\bf 65,825} & {\bf 0.228} & {\bf 0.008} \\
    \bottomrule
     \end{tabular}
    \caption{The total number of trainable parameters and some relevant outputs related to the QML models with Angle Encoding and different numbers of qubits in the PQC (specifically the best average RMS and the average STD calculated over four cross-validation folds with one hundred epochs).}
    \label{tab:qml_ang_comp}
    \end{table}    

From Table~\ref{tab:qml_ang_comp} and Figure~\ref{fig:QML_ang_7_14}, we can draw two key conclusions. First, the QML model with Angle Encoding does not demonstrate improved predictive performance with an increasing number of qubits. While there is a slight improvement with six and seven qubits, this effect saturates after eight qubits. Second, a comparison of Tables~\ref{tab:cml_comp} and~\ref{tab:qml_ang_comp} reveals that the QML model with Angle Encoding offers a slight improvement over the classical FNN. In all reported cases, the QML with Angle Encoding outperforms the classical FNN model in both stability (lower standard deviation) and effectiveness (lower average RMSE). 

As a final remark, Figure~\ref{fig:time_ang_amp} shows that the average execution time per epoch of the QML model with Angle Encoding increases exponentially with the number of qubits, consistent with the resource demands of the state-vector \texttt{default.qubit} simulator. For illustration purposes, in the same Figure~\ref{fig:time_ang_amp}, we also mark the execution time of the QML with Amplitude Encoding proposed in the current work, highlighting the slight advantage performance.

\subsection{Performance increasing the number of trees in XGBoost}
\label{app:xgb}

\begin{figure}
    \centering
    \includegraphics[scale=0.6]{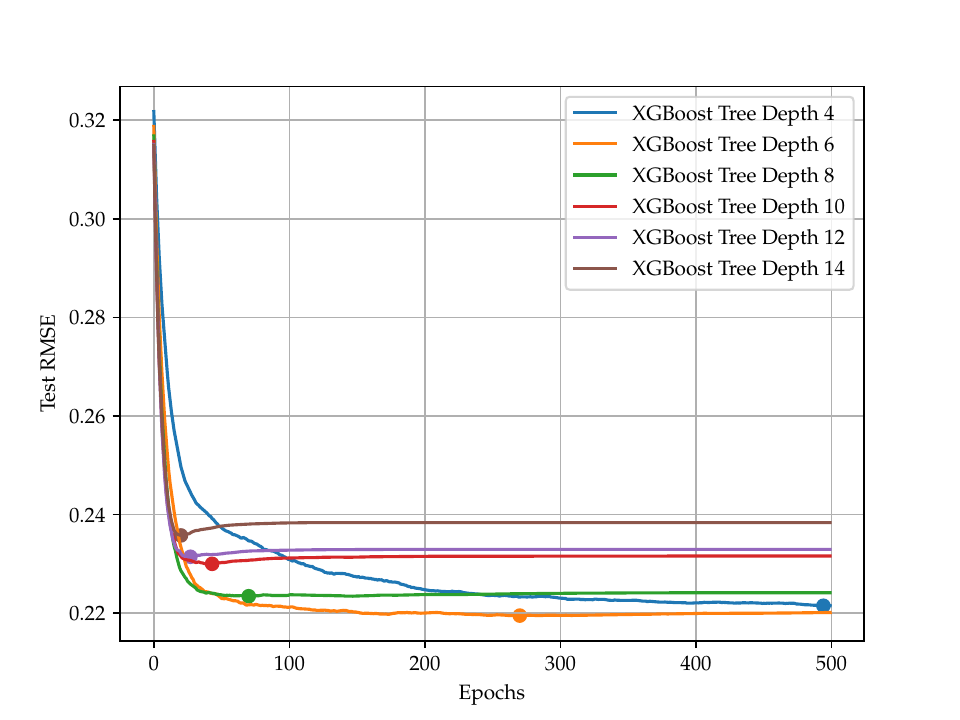}
    \caption{The average test RMSE calculated over four cross-validation folds of the XGBoost model with different tree depths. The dots indicate the corresponding epoch of the minimum RMSE achieved (which essentially indicates when the model converges).}
    \label{fig:trees_xgb}
\end{figure}

This Appendix illustrates a benchmark using XGBoost on the same regression task. We systematically varied the maximum tree depth parameter and recorded the corresponding test RMSE on a four-fold cross-validation, and the number of trees required to converge (Table~\ref{tab:xgb_trees} and Figure~\ref{fig:trees_xgb}).

\begin{table}[htbp]
\centering
\resizebox{1.\linewidth}{!}{
\begin{tabular}{p{2.3cm} p{1.3cm} p{2.cm} p{1.5cm} p{1.5cm} p{1.5cm}}
\toprule
Model & Max Tree depth & \# Parameters & Best Avg. RMSE & At Epoch/ Tree \# & Avg. STD  \\
\midrule
\multirow{11}{*}{XGBoost} &  4   & $\approx 8,000$  & 0.223 & 495 & 0.011\\
                          &  5   & $\approx 10,000$ & 0.223 & 305 & 0.015\\
                          &  6   & $\approx 18,000$ & {\bf 0.220} & 270 & 0.011 \\
                          &  7   & $\approx 22,000$ & 0.222 & 170 & 0.014\\ 
                          &  8   & $\approx 18,000$ & 0.224 & 70 & 0.014\\  
                          &  9   & $\approx 23,000$ & 0.227 & 45 & 0.014\\
\cmidrule(lr){2-6}
                          &  10  & $\approx 40,000$ & 0.230 & 43 & 0.015\\
                          &  11  & $\approx 81,000$ & 0.231 & 42 & 0.011\\
                          &  12  & $\approx 120,000$ & 0.232& 30 & 0.014\\
                          &  13  & $\approx 205,000$ & 0.235& 25 & 0.016\\
                          &  14  & $\approx 330,000$ & 0.236& 20 & 0.016\\
\hline
QML Amplitude Encoding & None & 65,825 & 0.228 & 55 & {\bf 0.008} \\
\bottomrule
\end{tabular}
}
    \caption{The total number of trainable parameters and key outputs of the XGBoost with different tree depths. The horizontal line indicates the cases where the XGBoost achieves a lower average RMSE (above the horizontal line) than our QML with Amplitude Encoding model.}
    \label{tab:xgb_trees}
\end{table}

Table~\ref{tab:xgb_trees} and Figure~\ref{fig:trees_xgb} offer several key findings. First, increasing tree depth leads to improved performance up to a point, specifically, depth six yields the lowest average RMSE. Beyond this, deeper trees result in no further accuracy gains and even degrade performance, likely due to overfitting. However, while shallower trees mitigate overfitting, they require a larger number of tree estimators to achieve convergence. For example, with depth six, XGBoost required around 270 trees. This increases model complexity and training time, reducing the overall efficiency of the approach. In addition, compared to XGBoost models with depth greater than nine, our QML model achieves better generalization performance (lower RMSE) while maintaining the same computational cost. Notably, the QML model also achieves the lowest standard deviation across test runs. This suggests that, despite XGBoost's lower RMSE with depth lower than nine, the difference may not be statistically significant due to the higher variability in XGBoost's results (see Section~\ref{sec:results_xgb}).

\section{Early stage results on QPU and Noisy Simulator}
\label{app:QPU}

\subsection{Preliminary results on real QPU Rigetti-Ankaa3 of the QML with Angle encoding}

In this appendix, we present preliminary results obtained using the proposed QML model on both a noisy quantum simulator and real quantum hardware. Before discussing the results, it is important to note that due to significant resource constraints, we trained our model on a relatively small dataset comprising 100 samples and 32 features, which we trained for just 13 epochs on the quantum hardware. This dataset was split into a single fold, with 75\% used for training and 25\% for testing.
 
Due to hardware compatibility limitations, only the QML model employing Angle Encoding was tested on a real quantum processor (see Figure~\ref{fig:ang_enc}). 

The real QPU used in this study was Rigetti’s Ankaa-3, a universal, gate-based quantum processor built with tunable superconducting qubits. Detailed specifications of the QPU can be found in~\cite{rigetti} and are summarized in Table~\ref{tab:rigetti}. 

For comparison purposes, we also implemented a noisy simulation of the quantum model. The noise model was designed based on the specifications of the Rigetti Ankaa-3 QPU~\cite{rigetti}, as outlined in Table~\ref{tab:rigetti}. The parameters used in the noisy simulation are reported in Table~\ref{tab:noisy_sim}.

\begin{table*}[htbp]
    \centering
    \resizebox{1.\linewidth}{!}{

    \begin{tabular}{p{2.5cm} p{2.5cm} p{2.5cm} p{2.5cm} p{2.5cm}}
    \toprule
    \multicolumn{5}{c}{{\bf Hardware Specifications}} \\
    \midrule
    Technology & $n$ Qubits & Native Gates & Gate Time & Gradient \\
    \midrule
    Superconducting qubits & 82 & $R_x$, $R_z$, ISWAP & $\SI{72}{ns}$  &  parameter-shift rule~\cite{schuld2019evaluating} \\   
    \bottomrule
    \multicolumn{5}{c}{{\bf Noise Specifications}} \\
    \hline
    Median T1 & Median T2 & Median f1QRB & Median fISWAP & Median fRO  \\
    \hline
    $\SI{31}{\mu s}$ & $\SI{19}{\mu s}$  & $99.91 \% \pm 0.01\%$  &  $98.58\%$ & $94.90\%$  \\   
    \bottomrule
    \end{tabular}}
    \caption{Specifications of the Rigetti Ankaa3 Quantum hardware~\cite{rigetti}. The T1 inidactes the relaxation time, so it indicates the time one qubit is able to keep is excited status corresponding to $\ket{1}$ before collapsing in the ground state $\ket{0}$. T2 inidactes the dephasing time, the time one qubit is able to keep is pure superposition states before, because of the interaction with the environment, to become a random mixture of quantum states. The Median f1QRB indicates the median fidelity of single-qubit gates measured using Randomized Benchmarking (RB), while the Median fISWAP for the two-qubit native gates. The Median fRO indicate the median fidelity due to read out operation of one qubit.}
    \label{tab:rigetti}
\end{table*}

\begin{table*}[htbp]
    \centering
    \resizebox{1.\linewidth}{!}{
    \begin{tabular}{p{2.5cm} p{2.5cm} p{2.5cm} p{2.5cm} p{2.5cm} p{2.5cm}}
    \toprule
    Prob. 1Q Depolarizing Channel & Prob. 2Q Depolarizing Channel & Prob. Amplitude Damping & Prob. Dephasing  & Prob. Read-Out Error\\
    \midrule
    0.09\% & 1.42\% & 0.023\% & 0.037\% & 5.1\% \\   
    \bottomrule
    \end{tabular}}
    \caption{Noisy parameter chosen to simulate the real QPU Rigetti Ankaa-3. These parameters are calculated accordingly to the value in Table~\ref{tab:rigetti}.}
    \label{tab:noisy_sim}
\end{table*}

\begin{figure}[htbp]
\begin{subfigure}{0.5\textwidth}
    \centering
        \makebox[\textwidth][l]{\hspace*{-1.in}\includegraphics[scale=.6]{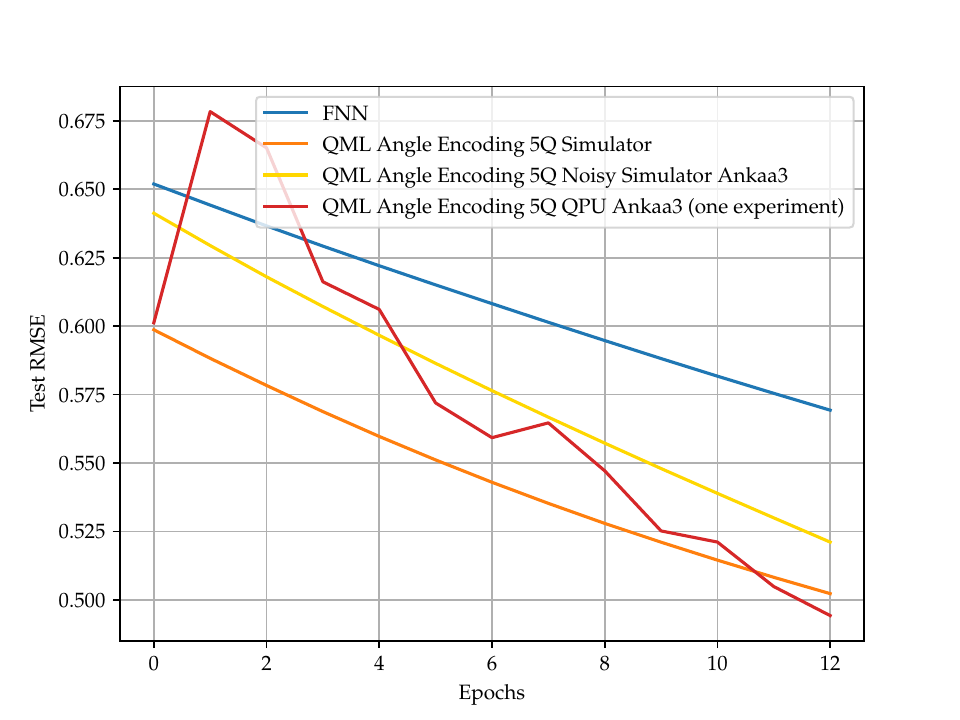}}
        \caption{Angle Encoding on Real QPU}
        \label{fig:qpu_noise}       
    \end{subfigure}
\begin{subfigure}{.5\textwidth}
    \centering
        \includegraphics[scale=.6]{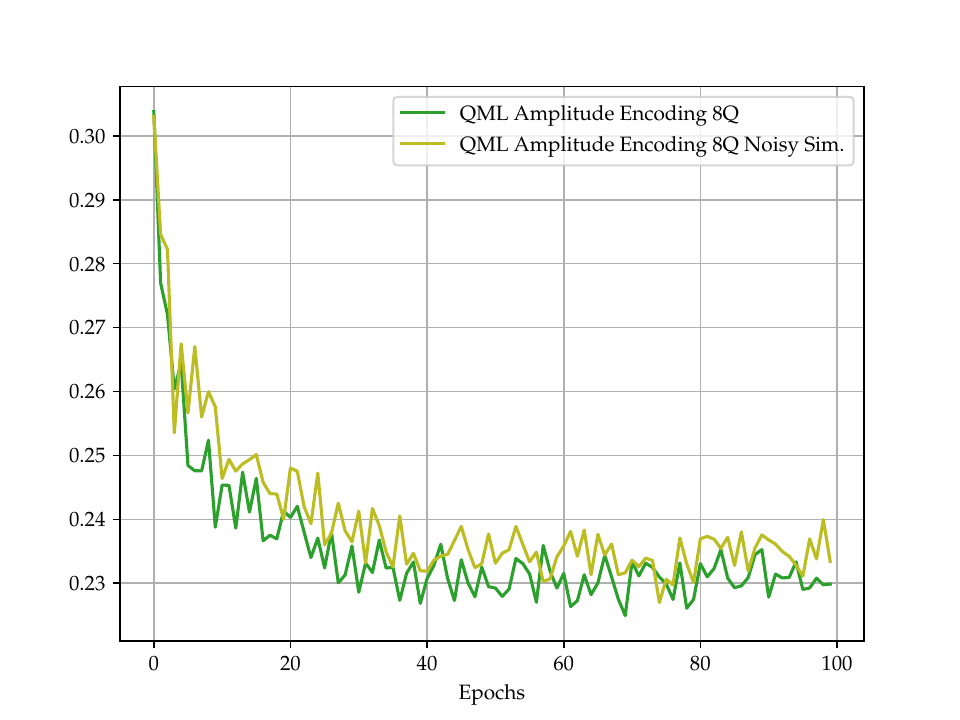}
        \caption{Amplitude Encoding on Noisy Simulator.}
        \label{fig:noise_amp}       
\end{subfigure}
\caption{(Left panel) The average test RMSE over ten different experiments of the FNN, the noisy QML model with Angle Encoding, and one single experiment on a real QPU. The models have been trained on a reduced dataset with one hundred observations and 32 features. (Right panel) The test RMSE of the noisy QML model with Amplitude Encoding, the noiseless corresponding model trained on the entire dataset.}
\end{figure}

Although our results are still preliminary and do not support a definitive conclusion, Figure~\ref{fig:qpu_noise} demonstrates the promising noise resilience of the QML model with Angle Encoding: during the first 13 epochs, both the noisy simulation and the execution on a real QPU outperform the classical FNN in terms of RMSE value.

\subsection{Preliminary results on noisy simulator of QML with Amplitude encoding}
Although the preliminary experiments on real hardware and noisy simulators provide early empirical evidence, it is equally important to assess the feasibility of deploying the proposed hybrid model on near-term quantum devices. We therefore summarize below several practical considerations related to qubit requirements, circuit depth, and the expected impact of device noise.

For the full model using Amplitude Encoding, a 256-dimensional input vector requires eight qubits, computed as the base-two logarithm of the feature dimension. This qubit count is within the physical availability of current mid-scale devices, such as those offered by IBM and IonQ. However, amplitude state preparation based on the~\cite{mottonen2004quantum} protocol (see also the supplementary materials provided) introduces a circuit depth that scales approximately linearly with the number of nonzero amplitudes. For dense 256-dimensional vectors, this can lead to circuits with several hundred single- and two-qubit operations. Current NISQ hardware typically supports reliable execution only up to a depth on the order of several tens to low hundreds of gates before coherence loss and accumulated noise significantly deteriorate performance. This constraint implies that exact amplitude loading, while theoretically possible, is challenging on today’s devices, leaving limited headroom for additional ancilla qubits, error-mitigation circuits, or a deeper variational layer. Angle Encoding, by contrast, requires far fewer gates and is therefore more compatible with existing NISQ hardware, as demonstrated in our Rigetti Ankaa-3 experiment.

Using the same noisy model described in Table~\ref{tab:noisy_sim}, we tested the full QML model with Amplitude Encoding, as shown in Figure~\ref{fig:ampl}. Due to the high computational cost of simulating noise on classical hardware, some limitations apply. We were able to train the QML with Amplitude Encoding on the full dataset for one hundred epochs, but only a single train/test split (75\%/25\%) was used at this stage.

Figure~\ref{fig:noise_amp} indicates that the QML model with Amplitude Encoding shows reasonable noise tolerance, with only minor deviations from the noiseless model, while the real-hardware run with Angle Encoding yields qualitatively similar behavior. Nevertheless, the limited scale of these experiments prevents any statistically significant assessment of the impact of hardware noise. Future work should therefore extend these tests to the full dataset, explore alternative encoding strategies, and conduct full training loops under realistic noise conditions to rigorously evaluate robustness and the potential need for noise-mitigation or error-correction protocols.

Several approaches exist to mitigate these limitations. Approximate or truncated state-preparation techniques can substantially reduce depth by loading only dominant components of the feature vector. Variational ansätze may also be compressed by removing redundant entangling layers. Standard noise-mitigation techniques (such as measurement-error correction, readout calibration, and zero-noise extrapolation) can further stabilize performance. These strategies outline a potential path toward running scaled-down versions of our model on near-term hardware.

These considerations indicate that while Angle Encoding is currently feasible on real devices, the full Amplitude Encoding architecture remains challenging for deployment on current NISQ hardware without circuit compression or approximation. The results shown in this appendix should therefore be viewed as early-stage feasibility tests rather than full demonstrations. These clarifications have been added to ensure a realistic and transparent assessment of the model’s implementability on existing quantum devices.

\end{appendices}

\clearpage
\pagestyle{empty}
\includepdf[pages=-,pagecommand={},offset=0 0]{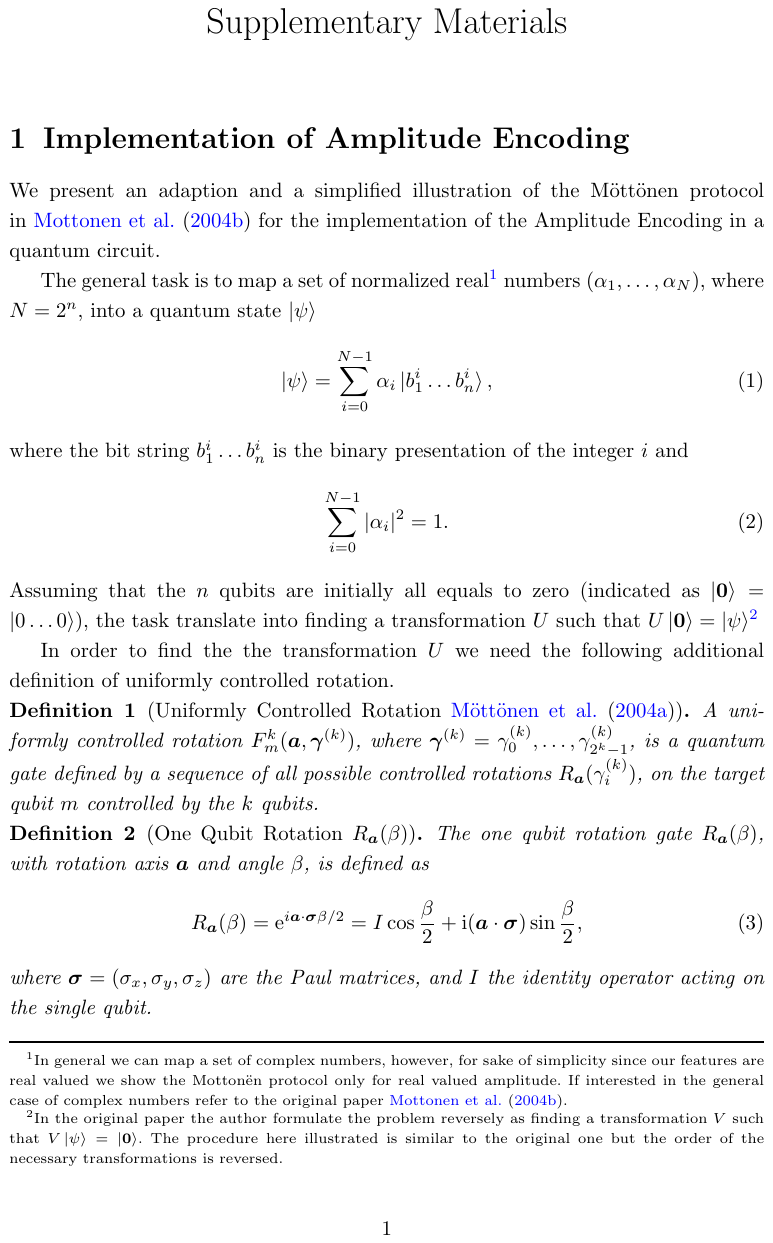}

\end{document}